%Paper: astro-ph/9310041
%From: rhb@het.brown.edu (Robert Brandenberger)
%Date: Sun, 24 Oct 93 09:24:36 EDT

\input phyzzx
\hfuzz=35pt
\FRONTPAGE
\line{\hfill BROWN-HET-906}
\line{\hfill May 1993}
\vskip1.0truein
\titlestyle{{TOPOLOGICAL DEFECTS AND STRUCTURE FORMATION}
\foot{Based on a Troisi\`eme Cycle lecture series given at the EPFL,
Lausanne, Switzerland, 4/22-5/13 1993.}}
\bigskip
\author{Robert H. Brandenberger}
\centerline{{\it Department of Physics}}
\centerline{{\it Brown University, Providence, RI 02912, USA}}
\bigskip
\abstract
Topological defects are produced during phase transitions in the very
early Universe. They arise in most unified theories of strong, weak and
electromagnetic interactions. These lectures focus on the role of
topological defects in cosmology, with particular emphasis on the models
of structure formation based on defects. The role of topological defects in
baryogenesis is also reviewed.
\endpage

\chapter{Introduction}
\par
Topological defects$^{1)}$ are inevitably produced during phase transitions
in the very early Universe. One or several types of defects arise in most
unified particle physics models of strong, weak and electromagnetic
interactions. Unless they are inflated away$^{2)}$ (i.e. unless there is a
period of inflation after the phase transition during which the defects
are produced), topological defects will play an important role in
cosmology (for recent reviews see e.g. Refs. 3-6). Main aim of these
lectures is to give a comprehensive pedagogical review of the role of
topological defects in cosmology.

The reason why topological defects can play a role in structure formation
in the early Universe is very simple. Topological defects carry energy.
This energy leads to an extra attractive gravitational force, and hence
the defects can act as seeds for cosmic structures. In particular, cosmic
strings (for a recent review see e.g. Ref.7) and global textures can lead
to attractive scenarios for the formation of galaxies and large-scale
structure.

In cosmology there are many open questions. What is the origin of the
observed large-scale structure in the Universe ?$^{9)}$ Why are galaxies
clustered? What explains the specific amplitudes and slopes of the galaxy
and galaxy cluster correlation functions?$^{10)}$ Do the models of structure
formation predict cosmic microwave anisotropies in agreement with the
recent COBE results$^{11)}$? There is a wealth of quantitative data in need of
a
consistent theoretical explanation.

Before 1980 there was no causal explanation starting from first principles
for any of the above questions. Around 1980 there was a crucial
realization which has now led to the development of several
self-consistent theories. The key point was to realize that in the very
early Universe, matter cannot be described in terms of an ideal gas. At
very high temperatures and energies, the matter content of the Universe
must be described in terms of quantum fields. In `old cosmology' (prior to
1980), space-time was described by general relativity and matter as an ideal
gas, in `new
cosmology' space-time is still described by general relativity, but matter in
terms
of fields. Evidently, from a conceptual point of view the present theories
are not satisfactory since they describe one sector in terms of classical
physics, the other quantum mechanically.

Since in `new cosmology' matter is described in terms of fields, the
possibility of having phase transition in the early Universe arises (for
current particle physics models of matter this is a rather generic
prediction). Hence, topological defects are formed, and they may play a
role in structure formation.

Concrete issues which must be addressed and which are discussed in these
lectures are:

- Why do defects form?

- What kind of defects arise?

- How to they evolve after formation?

- What are the mechanisms by which defects give rise to cosmic
structures?

- Are there direct observational signatures for topological defect
models?

Phase transition and therefore topological defects arise in many areas of
physics. In metals and other solids, defects arise during the fluid to
solid phase transition.$^{12)}$ The resulting crystal defects can be point,
line or planar defects. In liquid crystals, defects arise during the
transition from the disordered phase yo the ordered phase.$^{13)}$ There have
recently been some very ingenious studies of the dynamics of defect
formation in liquid crystals.$^{14)}$ Vortex lines also appear in the
transition from the normal to the superfluid phase of the $^{3}He$$^{15)}$, and
in the normal to superconducting transition in superconducting materials
$^{16)}$.

The above are all examples of topological defects in condensed matter
systems as opposed to relativistic field theory. There are important
similarities and differences between the two classes of systems. In
condensed matter systems the dynamics is generally friction dominated
and nonrelativistic, whereas in flat space-time the dynamics of
topological defects in field theory and cosmology is relativistic. Hence,
condensed matter systems give a good analogy for the topology and
formation of defects, but not for the dynamics.

These lecture notes begin with a survey of the successes and problems of
standard `big bang' cosmology. In Chapter 3, phase transitions and the
origin and nature of topological defects are reviewed. In Chapter 4, the
formation and evolution of topological defects (in particular cosmic
strings) in the early Universe are treated. Chapters 5-8 deal with
topological defect models for the formation of structure. First, an
introduction to the basic mechanisms and a survey of the most relevant
data is given. Then, the cosmic string and global texture models of
structure formation are developed, and in Chapter 8 the predictions of the
two models are compared with those of inflationary Universe scenarios.
Chapters 9 and 10 deal with the microphysics of cosmic strings. Baryon
number violating scattering processes based on strings are discussed in
Chapter 9, and the role of topological defects in baryogenesis is analyzed
in the final chapter. These lectures are based in part on lectures $^{17)}$
given at the $7^{th}$ Swieca Summer School in Particles and Fields. (see
also Ref. 18).

A word concerning notation. Unless mentioned otherwise, units in which $c
= \hbar = k_{B} = 1$ are employed. The metric $g_{\mu \nu}$ is taken to
have signature $(+,-,-,-)$. Greek indices ran over space and time, Latin
ones over spatial indices only. The Hubble expansion rate is $H(t)
=\dot{a} (t)/a(t)$, where $a(t)$ is the scale factor of a
Friedmann-Robertson-Walker (FRW) Universe. The present value of $H$ is $h
\times 100 km s^{-1} Mpc^{-1}$, where $0.4 < h <1$. Unless stated otherwise,
the value of $h$ is taken to be $0.5$. As usual, $G$ and $m_{pl}$ stand for
Newton's constant and Planck mass, respectively, and $z(t)$ is the
redshift at time $t$. Distances are measured in $Mpc$, where $1 Mpc$ is
$3.26 \cdot 10^6$ light years. $t_o$ usually denotes the present time, and
$T_0$ stands for the present temperature.

\chapter{Standard Cosmology:  Successes and Problems}

The standard big bang cosmology rests on three theoretical pillars:
the cosmological principle, Einstein's general theory of relativity
and a perfect fluid description of matter.

The cosmological principle$^{19)}$ states that on large distance
scales the Universe is homogeneous.  From an observational point of view
this is an extremely nontrivial statement.  On small scales the
Universe looks extremely inhomogeneous.  The inhomogeneities of the
solar system are obvious to everyone, and even by naked eye it is
apparent that stars are not randomly distributed.  They are bound into
galaxies, dynamical entities whose visible radius is about 10$^4$ pc.
Telescopic observations show that galaxies are not randomly
distributed, either.  Dense clumps of galaxies can be identified as
Abell clusters.  In turn, Abell cluster positions are correlated to
produce the large-scale structure dominated by sheets (or filaments)
with typical scale 100 Mpc observed in recent redshift
surveys$^{9)}$.  Until recently, every new survey probing the
Universe to greater depth revealed new structures on the scale of the
sample volume.  In terms of the visible distribution of matter there
was no evidence for large-scale homogeneity.  This situation changed
in 1992 with the announcement$^{20)}$ that a new redshift survey
complete to a depth of about 800 Mpc had discovered no prominent
structures on scales larger than 100 Mpc.  This is the first
observational evidence from optical measurements in favor of the
cosmological principle.  However, to put this result in perspective we
must keep in mind that the observed isotropy of the CMB temperature to
better than 10$^{-5}$ on large angular scales has been excellent
evidence for the validity of the cosmological principle.

The second theoretical pillar is general relativity, the theory which
determines the dynamics of the Universe.  According to general
relativity, space-time is a smooth manifold.  Together with the
cosmological principle this tells us that it is possible to choose a
family of hypersurfaces with maximal symmetry.  These are the
homogeneous constant time hypersurfaces.  The metric of these surfaces
is $^{21)}$
$$
ds^2 = a(t)^2 \, \left[ {dr^2\over{1-kr^2}} + r^2 (d\theta^2 + \sin^2
\theta \, d \varphi^2) \right] \eqno\eq
$$
when using spherical polar coordinates.  The constant $k$ is $+1$, $0$
or $-1$ for closed, flat or open surfaces respectively.  The function
$a(t)$ is the scale factor of the Universe.  By a coordinate choice,
it could be set equal to 1 at any given time.  However, the time
dependence of $a(t)$ indicates how the spatial sections evolve as a
function of time.  The full space-time metric is
$$
ds^2 = dt^2 - a (t)^2 \, \left[ {dr^2\over{1-kr^2}} + r^2 (d\theta^2 + \sin^2
\theta \, d \varphi^2) \right] \, . \eqno\eq
$$
According to the Einstein equivalence principle, test particles move
on geodesics with respect to the metric given by $ds^2$.  This implies
that the peculiar velocity $\underline{v}_p$ obeys the equation
$$
\underline{\dot v}_p + {\dot a\over a} \, \underline{v}_p = 0 \, . \eqno\eq
$$
Here,
$$
\underline{v}_p \equiv a (t) \, {d \underline{x}^c\over{dt}} \, , \eqno\eq
$$
$\underline{x}^c$ being the comoving coordinates $r$, $\theta$ and $\varphi$.
Equation (2.3) implies that in an expanding Universe, the peculiar
velocity $\underline{v}_p$ decreases:
$$
\underline{v}_p (t) \sim a^{-1} (t) \, . \eqno\eq
$$
Therefore, trajectories with constant $\underline{x}^c$ are geodesics and
correspond to particles at rest.  The velocity $\underline{v}_p$ is the
physical
velocity relative to the expansion of the Universe (see Fig. 1).
\goodbreak \midinsert \vskip 4.5cm
\hsize=6in \raggedright
\noindent {\bf Figure 1.} Sketch of the expanding Universe.
Concentric circles indicate space at fixed time, time increasing as
the radius gets larger.  Points at rest have constant comoving
coordinates.  Their world lines are straight lines through the origin
(e.g. $L$).
\endinsert

{}From observations$^{22)}$ it is known that the Universe is at present
expanding.  Looking at distant galaxies, we detect a redshift $z$ of
light which increases linearly with the distance $d$ of the
object:
$$
z (d) \sim d \, , \eqno\eq
$$
where the redshift $z$ is defined by
$$
z = {\lambda_0\over \lambda_e} - 1 \, , \eqno\eq
$$
with $\lambda_0$ and $\lambda_e$ being the wavelengths measured by us
and by the source.  The relationship (2.6) is explained by taking
galaxies to be at rest in comoving coordinates, and $a(t)$ to be
increasing.  In this case for $z \ll 1$
$$
z \simeq H (t_0) d \>\> \left[ H (t) = {\dot a \over a} (t) \right] \,
, \eqno\eq
$$
$t_0$ being the present time.

The most important consequence of general relativity for the history
of the Universe is that it relates the expansion rate to the matter
content.  The Einstein field equations follow from the action
$$
S = \int d^4 x \, \sqrt{-g} \, (R - 16 \pi G {\cal L}_M) \eqno\eq
$$
where $R$ is the Ricci scalar curvature, $g$ is the determinant of the
metric, and ${\cal L}_M$ is the Lagrange density for matter.
Evaluating the equations of motion obtained by varying (2.9) with
respect to $g_{\mu \nu}$ for a metric of the form (2.2) leads to the
famous Friedmann-Robertson-Walker (FRW) equations
$$
\left( {\dot a\over a} \right)^2 - {k\over a^2} = {8 \pi G\over 3} \,
\rho \eqno\eq
$$
$$
{\ddot a\over a} = - {4 \pi G\over 3} \> (\rho + 3p) \, . \eqno\eq
$$
Note that for fixed energy density $\rho$, the evolution of $a(t)$
depends in an important way on the pressure $p$.

Besides explaining Hubble's redshift-distance relation (2.6), standard
big bang cosmology makes two key quantitative predictions:  the
existence of a black body cosmic microwave background$^{23)}$ and
nucleosynthesis, the generation of light elements$^{23-25)}$.

Consider ordinary matter made up of atoms in an expanding Universe.
The energy density in matter scales as $a(t)^{-3}$, and the
temperature $T (t)$ as $a^{-1}(t)$.  Thus, standard cosmology predicts
that as we go back in time, the Universe was warmer.  In particular,
at a critical temperature $T_{rec}$, matter becomes ionized.  Before
$t_{rec}$ (the time corresponding to $T_{rec}$) the Universe was
opaque to photons, after $t_{rec}$ it was transparent.  To be more
precise, $T_{rec}$ is the temperature when photons fall out of thermal
equilibrium.  Thereafter they propagate without scattering.  The black
body nature of the spectrum of photons is maintained, but the
temperature redshifts.  Hence, the big bang model predicts$^{23,26)}$
a black body spectrum of photons of temperature
$$
T_0 = T_{rec} \, z^{-1}_{rec} \eqno\eq
$$
where the cosmological redshift $z(t)$ is given by
$$
1 + z (t) = {a (t_0)\over{a (t)}} \eqno\eq
$$
and $z_{rec} = z (t_{rec}) \sim 10^3$.

In 1965, Penzias and Wilson$^{27)}$ discovered this remnant black body
radiation at a temperature of about $3^\circ$ K.  Since the spectrum
peaks in the microwave region it is now called CMB (cosmic microwave
background).  Recent satellite (COBE)$^{28)}$ and rocket $^{29)}$
experiments have confirmed the black body nature of the CMB to very
high accuracy.  The temperature is 2.73$^\circ$ K $=T_0$.

Given the existence of the CMB, we know that matter has two
components: dust (with energy density $\rho_m (t)$) and radiation
(with density $\rho_r (t)$).  At the present time $t_0$, $\rho_m (t)
\gg \rho_r (t)$.  The radiation energy density is determined by $T_0$,
and the matter energy density can be estimated by analyzing the
dynamics of galaxies and clusters and using the virial theorem.  However, since
$$
\eqalign{
& \rho_m (t) \sim a (t)^{-3} \cr
& \rho_r (t) \sim a (t)^{-4} \, , } \eqno\eq
$$
as we go back in time the fraction of energy density in radiation
increases, and the two components become equal at $t_{eq}$, the time of
equal matter and radiation.  The corresponding redshift is
$$
z_{eq} = \Omega h^{-2}_{50} \, 10^4 \eqno\eq
$$
where
$$
\Omega = {\rho\over{\rho_{cr}}} (t_0) \, , \eqno\eq
$$
$\rho_{cr}$ being the density for a spatially flat Universe (the
critical density), and $h_{50}$ in the value of $H$ in units of 50 km
s$^{-1}$ Mpc$^{-1}$.

The time $t_{eq}$ is important for structure formation$^{30)}$.  It is
only after $t_{eq}$ that perturbations on scales smaller than the
Hubble radius $H^{-1} (t)$ can grow.  Before then, the radiation
pressure prevents growth.  A temperature-time plot of the early
Universe is sketched in Fig. 2.  Note that $t_{eq} < t_{rec}$.
\goodbreak \midinsert \vskip 6.5cm
\hsize=6in \raggedright
\noindent {\bf Figure 2.} Temperature-time diagram of standard big
bang cosmology.  The present time, time of decoupling and time of
equal matter and radiation are $t_0$, $t_{rec}$ and $t_{eq}$,
respectively.  The Universe is radiation dominated before $t_{eq}$
(Region A), and matter dominated in Region B.  Before and after
$t_{rec}$ the Universe is opaque and transparent, respectively, to
microwave photons.
\endinsert

The second quantitative prediction of standard big bang cosmology
concerns nucleosynthesis.  Above a temperature of about $10^{9 \circ}$
K, the nuclear interactions are sufficiently fast to prevent neutrons
and protons from fusing.  However, below that temperature, it is
thermodynamically favorable$^{23-25)}$ for neutrons and protons to
fuse and form deuterium, helium 3, helium 4 and lithium 7 through a
long and interconnected chain of reactions.  The resulting light
element abundances depend sensitively on the expansion rate of the
Universe and on $\Omega_B$, the fraction of energy density $\rho_B$ at
present in baryons relative to the critical density $\rho_c$.  In Fig.
3, recent$^{31)}$ theoretical calculations of the abundances are shown
and compared with observations.  Demanding agreement with all
abundances leaves only a narrow window
\goodbreak \midinsert \vskip 12.5cm
\hsize=6in \raggedright
\noindent {\bf Figure 3.} Light element abundances as a function of
the baryon to entropy ratio $\eta$.  $Y_p$ is the $^4$He mass
fraction, calculated assuming three light neutrino species, for two
different values $\tau_{1/2}$ of the neutrino half life (in minutes).
The abundances of $D + ^3$He and of $^7$Li are shown as ratios of
their number density relative to the number density of $H$.  The
horizontal lines indicate limits from observations: upper limit of
10$^{-4}$ for the $D + ^3$He abundance, and upper limits on $^7$Li
from observations of dwarf stars.  The $^7$Li curve is shown with $\pm
2 \sigma$ errors.  Combining $^4$He and $^3$He $+ D$ limits leaves
only a small window for $\eta$ which is allowed.  A major success of
primordial nucleosynthesis is that the $^7$Li abundance matches well.
Note that $\eta_{10}$ is $\eta$ in units of 10$^{-10}$.
\endinsert
$$
3 \times 10^{-10} < \eta < 10^{-9} \, , \eqno\eq
$$
where $\eta$ is the ratio of baryon number density $n_B$ to entropy
density $s$
$$
\eta = {n_B\over s} \, . \eqno\eq
$$
{}From (2.17), it follows that $\Omega_B$ is constrained:
$$
0.01 < \Omega_B h^2 < 0.035 \, . \eqno\eq
$$
In particular, if the Universe is spatially flat, there must be
nonbaryonic dark matter.  We will return to the dark matter issue
shortly.

In summary, the three observational pillars of standard big bang
cosmology are Hubble's redshift-distance relation, the existence and
black body spectrum of the CMB, and the concordance between observed
and theoretically determined light element abundances.

Standard big bang cosmology is faced with several important problems:
the age, dark matter, homogeneity, flatness, and formation of
structure problems (this list is not exhaustive$^{32)}$, but contains
what I consider to be the key problems).  In addition, standard big
bang cosmology does not explain the small value of the cosmological
constant -- a problem which no present cosmological model addresses
(see, however, Ref. 33 for some recent progress on this issue).  Of
the above five problems, only the first is a possible conflict of the
theory with observations.  The other four are questions which are left
unanswered by the theory.  Extensions of the standard model are needed
to address these issues.

Globular cluster ages have been estimated to lie in the range $13 - 18
\times 10^9$ years$^{34)}$.  Nuclear cosmochronology gives an age of
structures in the range $(10 - 20) \times 10^9$ yr.  However, big bang
cosmology (in the absence of a cosmological constant) predicts an age
$$
\tau = {7\over h} \cdot 10^2 {\rm yr} \, . \eqno\eq
$$
Thus, theory and observations are only compatible if $h < 0.55$.

The dark matter problem$^{35)}$ problem has various aspects.  There is
more matter in galaxies than is visible in stars.  This follows by
studying galactic velocity rotation curves and observing that the
velocity remains constant beyond the visible radius of the galaxy.
Whereas$^{34)}$ the contribution of stars to $\Omega$ is less than
$0.01$, galaxies as a whole contribute $ > 0.02$ to $\Omega$.  The
second level of the dark matter problem lies in clusters.  By studying
cluster dynamics it can be inferred that the contribution to $\Omega$
exceeds $0.1$.  Finally, observations on the largest scales give an
even larger contribution to $\Omega$.  From Virgo infall it follows
that $\Omega > 0.3$, and from large scale velocity measurements
(POTENT)$^{36)}$ or from infrared galaxy (IRAS) surveys$^{37)}$ it
follows that
$$
0.5 < \Omega < 3 \, . \eqno\eq
$$
Inflation predicts $\Omega \simeq
1$, unless fine tuned initial conditions are chosen$^{38)}$.

The bottom line of the dark matter problem is that more mass is
observed than can be in light (hence the name ``dark matter"), and
that most of this missing matter must be nonbaryonic (since by (2.19)
$\Omega_B < 0.14$).  Note that there must be some baryonic dark
matter, and that all of the galactic dark matter could be baryonic.
This comment will be relevant when discussing cosmic strings and
galaxy formation in Section 6.  If $h = 0.5$, then most of the cluster
dark matter must be nonbaryonic.  Standard cosmology does not address
the issue of what the dark matter is.  New cosmology can provide a
solution.  Quantum field theory models of matter in the Universe
give rise to several candidate particles which could constitute the dark
matter$^{39)}$.

The last three problems are the classic problems which the
inflationary Universe scenario addresses$^{2)}$.  In Fig. 4, the
homogeneity (or horizon) problem is illustrated.  As is sketched, the
region $\ell_p (t_{rec})$ over which the CMB is observed to be homogeneous
to better than one part in $10^5$ is much larger than the forward
light cone $\ell_f (t_{rec})$ at $t_{rec}$, which is the maximal
distance over which microphysical forces could have caused the
homogeneity:
$$
\ell_p (t_{rec}) = a(t_{rec}) \, \int\limits^{t_0}_{t_{rec}} \, dt \, a^{-1}
(t)
\simeq 3 t_0 \left( 1 - \left({t_{rec}\over t_0} \right)^{1/3} \right)
a(t_{rec})\eqno\eq
$$
$$
\ell_f (t_{rec}) = a (t_{rec}) \int\limits^{t_{rec}}_0 \, dt \, a^{-1}
(t) \simeq 3 t_0^{2/3} \, t^{1/3}_{rec} a(t_{rec})\, . \eqno\eq
$$
{}From the above equations it is obvious that $\ell_p (t_{eq}) \gg
\ell_f (t_{eq})$.  Hence, standard cosmology cannot explain the
observed isotropy of the CMB.
\goodbreak \midinsert \vskip 7.5cm
\hsize=6in \raggedright
\noindent {\bf Figure 4.} A sketch of the homogeneity problem: the
past light cone $\ell_p (t)$ at the time $t_{rec}$ of last scattering
is much larger than the forward light cone $\ell_f (t)$ at $t_{rec}$.
\endinsert

In standard cosmology and in an expanding Universe, $\Omega = 1$ is an
unstable fixed point.  This can be seen as follows.  For a spatially
flat Universe $(\Omega = 1)$
$$
H^2 = {8 \pi G\over 3} \, \rho_c \, , \eqno\eq
$$
whereas for a nonflat Universe
$$
H^2 + \varepsilon \, T^2 = \, {8 \pi G\over 3} \, \rho \, , \eqno\eq
$$
with
$$
\varepsilon = {k \over{(a T)^2}} \, . \eqno\eq
$$
The quantity $\varepsilon$ is proportional to $s^{-2/3}$, where $s$ is
the entropy density.  Hence, in standard cosmology, $\varepsilon$ is
constant.  Combining (2.24) and (2.25) gives
$$
{\rho - \rho_c\over \rho_c} = {3 \over{8 \pi G}} \> {\varepsilon T^2
\over \rho_c} \sim T^{-2} \, . \eqno\eq
$$
Thus, as the temperature decreases, $\Omega - 1$ increases.  In fact,
in order to explain the present small value of $\Omega - 1 \sim {\cal O}
(1)$, the initial energy density had to be extremely close to critical
density.  For example, at $T= 10^{15}$ GeV, (2.27) implies
$$
{\rho - \rho_c\over \rho_c} \sim 10^{-50} \, . \eqno\eq
$$
What is the origin of these fine tuned initial conditions?  This is
the flatness problem of standard cosmology.
\par
The last problem of the standard cosmological model I will mention is the
``formation of structure problem."  Observations indicate that galaxies and
even clusters of galaxies have nonrandom correlations on scales
larger than 50 Mpc$^{9,40,41)}$.  This scale is comparable to the comoving
horizon at $t_{eq}$. Thus,
if the initial density perturbations were produced much before $t_{eq}$, the
correlations cannot be explained by a causal mechanism.  Gravity alone is, in
general, too weak to build up correlations on the scale of clusters after
$t_{eq}$ (see, however, the explosion scenario of Ref. 42).  Hence, the two
questions of what generates the
primordial density perturbations and what causes the observed correlations, do
not have an answer in the context of standard cosmology.  This problem is
illustrated by Fig. 5.
\goodbreak\midinsert \vskip 7cm
\hsize=6in \raggedright
\noindent {\bf Figure 5.} A sketch (conformal separation vs. time) of the
formation of structure problem:  the comoving separation $d_c$ between two
clusters is larger than the forward light cone at time $t_{\rm rec}$.
\endinsert
\par
In 1981, based on previous work by many other people (see e.g., Refs.
43-45 for a
detailed bibliography of early work), Guth realized$^{2)}$ that having a
sufficiently long phase in the very early Universe during which the scale
factor expands exponentially
$$
a (t) \sim e^{Ht}\eqno\eq
$$
can potentially solve the three last problems listed above.  This phase of
exponential expansion is called the de Sitter or inflationary phase.
\topinsert \vskip 11cm
\hsize=6in \raggedright
\noindent {\bf Figure 6.} A sketch (physical coordinates vs. time) of the
solution of the homogeneity problem.  Due to exponential expansion the forward
light cone is larger than the past light cone at $t_{\rm rec}$.  The dashed
line is the apparent horizon or Hubble radius.
\endinsert
\par
In Fig. 6 it is sketched how a period of inflation can solve the homogeneity
problem.  $t_i$ shall denote the onset of inflation, $t_R$ the end.
$\triangle t = t_R - t_i$ is the period of inflation.  During inflation, the
forward light cone increases exponentially compared to a model without
inflation, whereas the past light cone is not affected for $t \ge t_R$.
Hence, provided $\triangle t$ is sufficiently large, $\ell_f (t_{rec})$ will be
greater than $\ell_p (t_{rec})$.  The condition on $\triangle t$ depends on the
temperature $T_R$ corresponding to time $t_R$, the temperature of reheating.
Demanding that $\ell_f (t_R) > \ell_p (t_R)$ we find, using (2.22) and
(2.23), the following criterion
$$
e^{\triangle t H} \ge \, {\ell_p(t_R)\over{\ell_f (t_R)}} \, \simeq \,
\left({t_0\over{t_R}}\right)^{1/2} \, = \, \left({T_R\over{T_0}}\right) \,
\sim \, 10^{27} \eqno\eq
$$
for $T_R \sim 10^{14}$GeV and $T_0 \sim 10^{-13}$GeV (the present microwave
background temperature).  Thus, in order to solve the homogeneity problem, a
period of inflation with
$$
\triangle t > 50 \, H^{-1}\eqno\eq
$$
is required.
\par
Inflation also can solve the flatness problem$^{2,46)}$.  The key point is
that the
entropy density $s$ is no longer constant.  As will be explained later, the
temperatures at $t_i$ and $t_R$ are essentially equal.  Hence, the entropy
increases during inflation by a factor $\exp (3 H \triangle t)$.  Thus,
$\epsilon$ decreases by a factor of $\exp (-2 H \triangle t)$.  With the
numbers used in (2.30):
$$
\epsilon_{\rm after} \sim 10^{-54} \epsilon_{\rm before}\, . \eqno\eq
$$
Hence, $(\rho - \rho_c)/\rho$ can be of order 1 both at $t_i$ and at the
present time.  In fact,if inflation occurs at all, the theory then predicts
that
at the present time $\Omega = 1$ to a high accuracy (now $\Omega < 1$ would
require special initial conditions).
\par
What was said above can be rephrased geometrically:  during inflation, the
curvature radius of the Universe -- measured on a fixed physical scale --
increases exponentially.  Thus, a piece of space looks essentially flat after
inflation even if it had measurable curvature before.
\par
Most importantly, inflation provides a mechanism which in a causal way
generates the primordial perturbations required for galaxies, clusters and
even larger objects$^{47-50)}$.  In inflationary Universe models, the
``apparent" horizon
$3t$ and the ``actual" horizon (the forward light cone) do not coincide at
late times.  Provided (2.30) is
satisfied, then (as sketched in Fig. 7) all scales within our apparent horizon
were inside the actual
horizon since $t_i$.  Thus, it is in principle possible to have a causal
generation mechanism for perturbations.

As will be shown at the beginning of the following chapter, in order to
obtain inflation it is necessary to describe matter in terms of field
theory. In particular, there must be a phase transition in the early
Universe during which the equation of state of matter changes, leading to
different expansion rates of the Universe before and after the
transition.

In order to obtain sufficient inflation $^{2)}$, the phase transition must be
slow. For particle physics motivated matter theories this is hard to
achieve as will again be shown in the next chapter. However, matter
theories generically predict phase transitions, and the latter lead to the
formation of topological defects.

Topological defect models provide a second mechanism for producing
structures in the Universe on large scales. To be specific, consider a
model giving rise to cosmic strings (see Chapter 4). The strings produced
during the phase transition will form a network of random walks. This
network has infinite extent and gives rise to nonrandom correlations on
all scales for point objects seeded by the strings (see Chapter 6). This
and similar topological defect models for structure formation will be the
focus of a substantial part of these lectures.
\goodbreak \midinsert \vskip 10.5cm
\hsize=6in \raggedright
\noindent {\bf Figure 7.} A sketch (physical coordinates vs. time) of the
solution of the formation of structure problem.  The separation $d_c$ between
two clusters is always smaller than the forward light cone.  The dashed line
is the Hubble radius $H^{-1} (t)$.
\endinsert
\par
\chapter{Phase Transitions and Topological Defects}
\section{Scalar Fields and Cosmology}
\par
Let us for a moment return to the inflationary Universe. As stated in
(2.29), the requirement for inflation is to have a time interval during
which the scale factor $a(t)$ expands exponentially. Recall form (2.10)
and (2.11) that for a spatially flat FRW Universe the Einstein equations
reduce to
$$\left({\dot{a}\over a}\right) ^2 = {8\pi G\over 3}\rho \eqno\eq $$
and
$$2{\ddot{a}\over a}+ \left({\dot{a}\over a}\right) ^2 = -8\pi G p\eqno\eq$$
However, using $a(t)=e^{tH}$, the left hand side of (3.2) can be evaluated
using (3.1) to give
$$2{\ddot{a}\over a} + \left({\dot{a}\over a}\right) ^2 = 8\pi G\rho.\eqno\eq
$$
In order that (3.2) and (3.3) agree, the equation of state of matter must
be
$$p=-\rho\eqno\eq $$
Therefore, 'new cosmology' (i.e. description of matter in terms of
fields) is required in order to obtain inflation.

It is not hard to show that an equation of state like (3.4) with negative
pressure can be obtained if matter is described in terms of scalar fields.
Consider the Lagrangian $L(\varphi)$ for a theory of a scalar field
$\varphi(\underline{x},t)$:
$${\cal L}(\varphi ) = {1\over 2}\partial_\mu\varphi\partial^\mu\varphi -
V(\varphi ) .\eqno\eq $$
Given the Lagrangian, the energy-momentum tensor $T_{\mu\nu}$ can be
determined as in any classical field theory (see eq. Ref. 51). In a
Universe with FRW metric
$$g_{\mu\nu} = diag\left(1,-a^2(t),-a^2(t),-a^2(t)\right) \eqno\eq $$
we obtain for  $\rho=T_{00}$ and $p={1\over 3} \sum_{i=1}^{3} T_{ii}$:
$$
\eqalign{\rho (\undertext{x}, t) & = \, {1\over 2} \dot \varphi^2 \,
(\undertext{x}, t) + \, {1\over 2} a^{-2} \, (\bigtriangledown \varphi)^2 +
V(\varphi)\cr
p (\undertext{x}, t) & = \, {1\over 3} \sum^3_{i = 1} T_{ii} \, = \, {1\over
2} \dot \varphi^2 \, (\undertext{x}, t) - {1\over 6} \, a^{-2}
(\bigtriangledown \varphi)^2 - V (\varphi)\, .}\eqno\eq
$$
Thus, if $\varphi (\undertext{x}, t_i) =$const
and $\dot \varphi (\undertext{x}, t_i) = 0$ at some initial time $t_i$
and $V (\varphi (\undertext{x}, t_i)) > 0$,
then the equation of state becomes $p = - \rho$ and leads to inflation.
\par
\midinsert \vskip 5cm
\hsize=6in \raggedright
\noindent {\bf Figure 8.}  A sketch of two potentials which can give rise to
inflation.
\endinsert
Two examples which give inflation are shown in Fig. 8.  In (a), inflation
occurs at the stable fixed point
$\varphi (\undertext{x}, t_i) = 0 = \dot \varphi (\undertext{x}, t_i)$.
However, this model is ruled out by observation: the inflationary
phase has no ending.  $V(0)$ acts as a permanent nonvanishing cosmological
constant.  In (b), a finite period of inflation can arise if
$\varphi(\undertext{x})$ is trapped at the local minimum $\varphi = 0$ with
$\dot \varphi (\undertext{x}) = 0$.  However, in this case $\varphi
(\undertext{x})$ can make a sudden transition at some time $t_R > t_i$ through
the potential barrier and move to $\varphi (\undertext{x}) = a$.  Thus, for
$t_i < t < t_R$ the Universe expands exponentially, whereas for $t > t_R$ the
contribution of $\varphi$ to the expansion of the Universe vanishes and we get
the usual FRW cosmology.  There are two obvious questions: how does the
transition occur and why should the scalar field have $V(\varphi) = 0$ at the
global minimum.  In the following section the first question will be
addressed.  The second question is part of the cosmological constant problem
for which there is as yet no convincing explanation.  Before studying
the dynamics of the phase transition, we need to digress and discuss
finite temperature effects.
\section{Finite Temperature Field Theory}
\par
The evolution of particles in vacuum and in a thermal bath are very different.
Similarly, the evolution of fields changes when coupled to a thermal bath.
Under certain conditions, the changes may be absorbed in a temperature
dependent potential, the finite temperature effective
potential$^{52)}$.
Here, a heuristic derivation of this potential will be given. The reader is
referred to Ref.53 or to the original articles$^{52)}$ for the actual
derivation.

        We assume that the scalar field $\varphi (\undertext{x},t)$ is
coupled to a thermal bath which is represented by a second scalar field
$\psi (\undertext{x}, t)$ in thermal equilibrium. The
Lagrangian for $\varphi$ is
$$
{\cal L} = {1\over 2} \partial_\mu \varphi \partial^\mu \varphi \, - \,
V(\varphi) \, - \, {1\over 2} \hat \lambda \varphi^2 \psi^2\, , \eqno\eq
$$
where $\hat \lambda$ is a coupling constant.  The action from which the
equations of motion are derived is
$$
S \, = \, \int d^4 x \, \sqrt{-g} {\cal L}\eqno\eq
$$
where $g$ is the determinant of the metric (3.6).   The resulting equation of
motion for $\varphi (\undertext{x},t)$ is
$$
\ddot \varphi + 3 H \dot \varphi \, - a^{-2} \bigtriangledown^2 \varphi \, =
\,- V^\prime (\varphi) - \hat \lambda \psi^2 \varphi \, . \eqno\eq
$$
If $\psi$ is in thermal equilibrium, we may replace $\psi^2$ by its thermal
expectation value $<\psi^2>_T$.  Now,
$$
<\psi^2>_T \sim T^2\eqno\eq
$$
which can be seen as follows:  in thermal equilibrium, the energy density of
$\psi$ equals that of one degree of freedom in the thermal bath.  In
particular, the potential energy density $V (\psi)$ of $\psi$ is of that order
of magnitude.  Let
$$
V (\psi) \, = \, \lambda_\psi \psi^4\eqno\eq
$$
with a coupling constant $\lambda_\psi$ which we take to be of the order 1
(if $\lambda_\psi$ is too small, $\psi$ will not be in thermal equilibrium).
Since the thermal energy density is proportional to $T^4$, (3.11) follows.
(3.10) can be rewritten as
$$
\ddot \psi + 3H \dot \varphi \, - \, a^{-2} \bigtriangledown^2 \varphi \, = \,
- V_T^\prime (\varphi),\eqno\eq
$$
where
$$
V_T (\varphi) \, = \, V (\varphi) \, + \, {1\over 2} \hat \lambda T^2
\varphi^2\eqno\eq
$$
is called the finite temperature effective potential.  Note that in
(3.14),
$\hat \lambda$ has been rescaled to absorb the constant of proportionality in
(3.11).
\par
These considerations will now be applied to Example A, a scalar field model
with
potential
$$
V (\varphi) \, = \, {1\over 4} \lambda (\varphi^2 - \eta^2)^2\eqno\eq
$$
($\eta$ is called the scale of symmetry breaking).  The finite temperature
effective potential becomes (see Fig. 9)
$$
V_T (\varphi) \, = \, {1\over 4} \lambda \varphi^4 - {1\over 2}
\, \left(\lambda \eta^2 - \hat \lambda T^2\right) \varphi^2 +
\, {1\over 4} \, \lambda \eta^4 \, . \eqno\eq
$$
For very high temperatures, the effective mass term is positive
and hence the energetically favorable state is $<\varphi> = 0$.
For very low temperatures, on the other hand, the mass term has a
negative sign which leads to spontaneous symmetry breaking.  The
temperature at which the mass term vanishes defines the critical
temperature $T_c$
$$
T_c \, = \, \hat \lambda^{-1/2} \lambda^{1/2} \eta\,.\eqno\eq
$$
\midinsert \vskip 6cm
\hsize=6in \raggedright
\noindent {\bf Figure 9.}  The finite temperature effective potential
for Example A.
\endinsert
\hfuzz=15pt
\par
As Example B, consider a theory with potential
$$
 V (\varphi) = \, {1\over 4} \varphi^4 - {1\over 3} \, (a + b) \varphi^3 +
{1\over 2} ab \varphi^2\eqno\eq
$$
with ${1\over 2} a > b > 0$.  The finite temperature effective potential is
obtained by adding ${1\over 2} \hat \lambda  T^2 \varphi^2$ to the
right hand side of (3.18).  ${ V_T} (\varphi)$
is sketched in Fig. 10 for various values of $ T$.  The critical temperature
$T_c$ is defined as the temperature when the two minima of ${V_T}(\varphi)$
become degenerate.
\par
\midinsert \vskip 6cm
\hsize=6in \raggedright
\noindent {\bf Figure 10.} The finite temperature effective potential
for Example B.
\endinsert
It is important to note that the use of finite temperature effective potential
methods is only legitimate if the system is in thermal equilibrium.  This
point was stressed in Refs. 54 and 55, although the fact should be
obvious from the derivation given above.  To be more precise, we require the
$\psi$ field to be in thermal equilibrium and the coupling constant $\hat
\lambda$ of (3.8) which mediates the energy exchange between the $\varphi$
and $\psi$ fields to be large.  However, (see e.g. Ref. 18), observational
constraints stemming from the amplitude of the primordial energy density
fluctuation spectrum force the self coupling constant $\lambda$ of $\varphi$
to be extremely small.  Since at one loop order, the interaction term ${1\over
2} \hat \lambda \varphi^2 \psi^2$ induces contributions to $\lambda$, it is
unnatural to have $\lambda$ very small and $\hat \lambda$ unsuppressed.
Hence, in many inflationary Universe models - in particular in new
inflation$^{56)}$ and in chaotic inflation$^{54)}$ - finite temperature
effective potential methods are inapplicable.
\section{Phase Transitions}
\par
The temperature dependence of the finite temperature effective potential in
quantum field theory leads to phase transitions in the very early Universe.
These transitions are either first or second order.
\par
Example $A$ of the previous section provides a model in which the transition
is second order (see Fig. 9).  For $T \gg T_c$, the expectation value of the
scalar field $\varphi$ vanishes at all points $\undertext{x}$ in space:
$$
< \varphi (\undertext{x}) > = 0 \, . \eqno\eq
$$
For $T < T_c$, this value of $< \varphi (\undertext{x}) >$ becomes unstable
and $< \varphi (\undertext{x})>$ evolves smoothly in time to a new value $\pm
\eta$.  The direction is determined by thermal and quantum fluctuations and
is therefore not uniform in space.  There will be domains of average radius
$\xi (t)$ in which $< \varphi (\undertext{x}) >$ is coherent.  By causality,
the coherence length is bounded from above by the horizon.  However, typical
values of $\xi (t)$ are proportional to $\lambda^{-1} \eta^{-1}$ if
$\varphi$ was in thermal equilibrium before the phase
transition$^{1)}$.
\par
In condensed matter physics, a transition of the above type is said to proceed
by spinodal decomposition$^{57)}$, triggered by a rapid quench.
\par
In Example B of the previous section, (see Fig. 10) the phase
transition is first order.  For $T > T_c$, the expectation value
$<\varphi (x) >$ is approximately $0$, the minimum of the high
temperature effective potential.  Provided the zero temperature
potential has a sufficiently high barrier separating the metastable
state $\varphi = 0$ from the global minimum (compared to the energy
density in thermal fluctuations at $T = T_c$), then $\varphi
(\undertext{x})$ will remain trapped at $\varphi = 0$ also for $T <
T_c$.  In the notation of Ref. 58, the field $\varphi$ is trapped in
the false vacuum.  After some time (determined again by the potential
barrier), the false vacuum will decay by quantum tunnelling.
\par
Tunnelling in quantum field theory was discussed in Refs. 58-61 (for
reviews see e.g., Refs. 62 and 53).  The transition proceeds by bubble
nucleation.  There is a probability per unit time and volume that at a
point $\undertext{x}$ in space a bubble of ``true vacuum" $\varphi
(\undertext{x}) = a$ will nucleate.  The nucleation radius is
microscopical.  As long as the potential barrier is large, the bubble
radius will increase with the speed of light after nucleation.  Thus,
a bubble of $\varphi = a$ expands in a surrounding ``sea" of false
vacuum $\varphi = 0$.
\par
To conclude, let us stress the most important differences between the
two types of phase transitions discussed above. In a second order
transition, the dynamics is determined mainly by
classical physics.  The transition occurs homogeneously in space
(apart from the phase boundaries which -- as discussed below -- become
topological defects), and $< \varphi (x) >$ evolves continuously in
time.  In first order transitions, quantum mechanics is essential.
The process is extremely inhomogeneous, and $< \varphi (x) >$ is
discontinuous as a function of time.

\section{Particle physics connection}
\par
Up to this point in these lectures, I have only discussed scalar field toy
models. The actual case of interest is a unified gauge theory of strong,
weak and electromagnetic interactions. Such a theory (see e.g. Ref. 63)
contains massless fermion fields $\psi$, gauge fields $A_\mu$ and scalar
fields $\varphi$. The Lagrangian $L(\psi,A_{\mu},\varphi)$ is invariant
under the action of some internal symmetry group $G$. This grand unified
symmetry is broken spontaneously (see e.g. Ref. 64) in one or several
stages for the `standard model' symmetry group $SU(3)\times SU(2)\times
U(1)$:
$$G\rightarrow H\rightarrow\dots\rightarrow SU(3)\times SU(2)\times U(1).
\eqno\eq $$
Spontaneous symmetry breaking is achieved by postulating that a scalar
(Higgs) field takes on an expectation value which is not invariant under
the full group $G$, but only under the subgroup $H$. In order that this be
possible, the potential $V(\varphi)$ for this Higgs field $\varphi$ must
have a global minimum at $\varphi \neq 0$ but a high temperature minimum at
$\varphi = 0$. Therefore, such a model will lead to a symmetry breaking
phase transition in the early Universe. Hence, the possibility of
topological defect formation arises.

As a toy model, consider a theory of a two component real scalar field
$$\varphi = \left (\matrix{
\varphi_1\cr
\varphi_2\cr
}\right ), \eqno\eq $$
with Lagrangian
$${\cal L}(\varphi ) = {1\over 2}\partial _\mu\varphi\partial ^\mu\varphi -
V(\varphi ), \eqno\eq $$
and symmetry breaking potential
$$V(\varphi ) = {1\over 4}\lambda (\varphi ^2 - \eta^2 )^2. \eqno\eq $$
Summation over the internal index $i=1,2$ is implicit in (3.22) and
(3.23). Obviously, the Lagrangian is invariant under a $U(1)$  rotation
$U(\alpha)$ in an internal field space $\Re^2$. However, a choice of vacuum
state
$$\varphi_{vac} = \eta\left (\matrix{
1\cr
0\cr
}\right ) \eqno\eq $$
breaks this symmetry completely.

The concept of vacuum manifold $\cal M$ will be crucial in the following
sections.  In general, $\cal M$ is defined  is the set of field
configurations which minimize the free energy modular gauge
transformations. For a gauge theory with symmetry breaking scheme as in
the first stage of (3.20)
$${\cal M}\simeq G/H . \eqno\eq $$
In the above toy model,
$${\cal M} = \{\varphi : V(\varphi )=V_{min} = 0 \} , \eqno\eq $$
I will now argue that in `realistic' particle physics models, a symmetry
breaking phase transition proceeds too fast to allow for inflation. Hence,
the topological defects which form in the phase transition are not
inflated away (i.e. moved beyond the present Hubble radius $H^{-1}(t)$ by
the  exponential expansion of the forward light cone), but remain inside
the Hubble radius and  are therefore important for cosmology.

The argument is based on the equation  of motion for the Higgs field.
Neglecting inhomogeneitis in $\varphi$, interactions with other fields,
and the expansion of the Universe, this equation becomes
$$\ddot{\varphi }\simeq - {\partial V\over \partial\varphi}\simeq
\lambda\eta^2\varphi \eqno\eq $$
The typical time scale $\tau$  for the solution is
$$\tau = \lambda^{-1/2}\eta^{-1} . \eqno\eq $$
Sufficient inflation requires
$$\tau \gg H^{-1} . \eqno\eq $$
However, from the FRW equation (3.1) it follows that
$$H^2\simeq {8\pi G\over 3}V(0) = {2\pi\over 3}\lambda G\eta^4 , \eqno\eq $$
and hence
$$\tau\ll H^{-1} \eqno\eq $$
unless $\eta > m_{pl}$.

In order to obtain inflation, it is in general $^{54,65)}$ necessary to
introduce a scalar field $\varphi$ which is {\it not} the Higgs field of a
unified field theory. We shall not further consider such models and
restrict our attention to matter theories motivated by particle
physics

\section{Classification of Topological Defects}\
\par
Different particle physics models admit different types of topological
defects. The topology of the vacuum manifold $\cal M$ determines the type
of defect.

The classification of defects is based on homotopy classes of $\cal M$. To
introduce the ideas (see e.g. Ref. 66), consider maps $\psi$ from $S^n$
to $\cal M$, $n$ being an integer. Two maps $\psi_1$ and $\psi_2$ are called
homotopically equivalent $(\psi_{1}\sim \psi_{2})$ if there exists a
continuous one parameter family of maps
$$\psi (t) : S^n\rightarrow {\cal M}\;\;\;\;\; t\in [0,1] \eqno\eq $$
with
$$\psi (0) = \psi_1\; ,\;\psi(1) = \psi_2 .\eqno\eq $$
The n'th homotopy `group' of $\cal M$, $\Pi_{n} ({\cal M})$, is the set of
all homotopy classes of maps $S^n \rightarrow {\cal M}$. Except for
$n = 0, \Pi_{n}({\cal M})$ is a group.

Let us consider a symmetry breaking pattern
$$ G\rightarrow H\eqno\eq $$
of a group $G$ with
$$\Pi_o (G) = \Pi_1 (G) = 1 \eqno\eq $$
to a subgroup $H$. As stated in (3.25), the vacuum manifold $\cal M$ is
isomorphic to $G/H$. Using exact sequences $^{66)}$ it can be shown $^{1,4)}$
that
$$\Pi_1({\cal M}) = \Pi_o (H) \eqno\eq $$
and
$$\Pi_2 ({\cal M}) = \Pi_1 (H) ,\eqno\eq $$
assuming $\Pi_{2} (G) = 1$  for  the second relation to hold.

Let us consider two relevant examples. The smallest grand unified group is
$SU(5)$. If we consider the symmetry breaking pattern
$$SU(5)\rightarrow SU(3)\times SU(2)\times U(1) , \eqno\eq $$
we can use (3.36) and (3.37) to conclude that
$$\Pi_1 ({\cal M}) = 1 \eqno \eq $$
and
$$\Pi_2 ({\cal M}) = \Pi_1 (U(1)) = {\cal Z} . \eqno\eq $$
As will be shown in the following section, (3.39) and (3.40) imply that no
cosmic strings but monopoles form during the phase transition associated
with (3.38).

As a second example, consider $G = SO(10)$ and the symmetry breaking
pattern
$$SO(10)\rightarrow SU(5)\times{\cal Z }_2 . \eqno\eq $$
In this example
$$\Pi_1 ({\cal M}) = {\cal Z}_2 , \eqno\eq $$
and hence cosmic strings form during the symmetry breaking phase
transition.

Let us give a brief overview of the classification of topological defects.
To be specific, consider a theory with an n-component real scalar field
$\varphi$ with symmetry breaking potential (3.23).

There are various types of local and global topological defects$^{3)}$
(regions of trapped energy density) depending on the number of components of
$\varphi$.  The words ``local" and ``global" refer to whether the symmetry
which is broken is a gauge or global symmetry.  In the case of local
symmetries, the topological defects have a well defined core outside of which
$\varphi$ contains no energy density in spite of nonvanishing gradients
$\nabla \varphi$:  the gauge fields $A_\mu$ can absorb the gradient,
{i.e.,} $D_\mu \varphi = 0$ when $\partial_\mu \varphi \neq 0$,
where the covariant derivative $D_\mu$ is defined by
$$
D_\mu = \partial_\mu - ig \, A_\mu \, , \eqno\eq
$$
$g$ being the gauge coupling constant.
Global topological defects, however, have long range density fields and
forces.
\par
Table 1 contains a list of topological defects with their topological
characteristic.  A ``v" marks acceptable theories, a ``x" theories which are
in conflict with observations (for $\eta \sim 10^{16}$ GeV).
\vskip.5cm
\vskip4.5cm

%\input /LocalApps/InstantTex/TeXTables.app/stables.tex
%\begintable
%{\bf Table 1} | $n$ | topology | local defect | global defect \elt
% domain wall \hfill | 1 | $\Pi_0 ({\cal M}) \neq 1$ | x | x \elt
% cosmic string \hfill | 2 | $\Pi_1 ({\cal M}) \neq 1$ | v | v \elt
% monopole \hfill | 3 | $\Pi_2 ({\cal M}) \neq 1$ | x | v \elt
% texture \hfill | 4 | $\Pi_3 ({\cal M}) \neq 1$ | - | v
%\endtable
%\catcode`\|=12

%\table
%{\bf Table 1} | $n$ | topology | local defect | global defect \cr
%domain wall \hfill | 1 | $\Pi_0 ({\cal M}) \neq 1$ | x | x \nr
%cosmic string \hfill | 2 | $\Pi_1 ({\cal M}) \neq 1$ | v | v \nr
%monopole \hfill | 3 | $\Pi_2 ({\cal M}) \neq 1$ | x | v \nr
%texture \hfill | 4 | $\Pi_3 ({\cal M}) \neq 1$ | - | v \endtable
%\vbox{\offinterlineskip
%\vskip.5cm
\par

In the following sections of this chapter we will construct the various
types of defects. The cosmological implications will be discussed in later
chapters.

The easiest defect to construct is the domain wall. Consider $n  = 1$ (or
more generally  a theory with $\Pi_{0} ({\cal M}) \neq 1$). In this case,
the vacuum manifold consists of two points
$$ \varphi_{vac} = \pm\eta .\eqno\eq $$
During the symmetry breaking phase transition, regions in physical space
${\cal R}^3$ with $\varphi = \pm \eta$ will form. These regions are separated
by
two dimensional surfaces (walls) with $\varphi \not\in {\cal M}$ (see Fig.
11). These are the domain walls. Since $\varphi\not \in {\cal M}$ in the
walls, $V (\varphi) > 0$ and hence the walls carry energy per unit area.
Via the usual gravitational force, this energy can act as a seed for
structures in the Universe.
\goodbreak \midinsert \vskip 4.5cm
\hsize=6in \raggedright
\noindent {\bf Figure 11.}
A two dimensional cross section through space showing a domain wall (DW)
separating a region with $\varphi =\eta $ ($+$) from a neighboring region with
$\varphi =-\eta $ ($-$).
\endinsert

\section{Cosmic Strings}
\par
Consider a theory in which matter consists of a gauge field $A_\mu$ and a
complex scalar field $\phi$ whose dynamics is given by the Lagrangian
$$
{\cal L} = {1\over 2} \, D_\mu \phi D^\mu \phi - V (\phi) + {1\over 4} \,
F_{\mu \nu} \, F^{\mu \nu}  \eqno\eq
$$
where $F_{\mu \nu}$ is the field strength tensor.
The potential $V(\phi)$ has the symmetry breaking ``Mexican hat" shape (see
Figure 12):
$$
V (\phi) = {1\over 4} \lambda (|\phi|^2 - \eta^2)^2 \, . \eqno\eq
$$
Hence, the vacuum manifold ${\cal M}$, the space of minimum energy density
configurations, is a circle $S^1$.
\midinsert \vskip 6.5cm
\hsize=6in \raggedbottom
\noindent{\bf Figure 12.} The zero temperature potential energy of the
complex scalar field used in the cosmic string model.
\endinsert
\par
The theory described by (3.45) and (3.46) admits one dimensional topological
defects, cosmic strings.  In the Abelian Higgs model of this example
the string solutions were first found by Nielsen and Olesen$^{67)}$.
It is possible to construct string
configurations which are translationally invariant along the $z$ axis.  On a
circle $C$ in the $x-y$ plane with radius $r$(see Fig. 13), the boundary
conditions
for $\phi$ are
$$
\phi (r, \theta) = \eta \, e^{i \theta} \, \eqno\eq
$$
where $\theta$ is the polar angle along $C$.
\goodbreak \midinsert \vskip 4.5cm
\hsize=6in \raggedright
\noindent {\bf Figure 13.}
Sketch of the cosmic string construction of section 3.6 . See text for
notation.
\endinsert
The
configuration (3.47) has winding number 1: at all points of the circle, $\phi$
takes on values in ${\cal M}$, and as $\varphi$ varies from 0 to $2 \pi$,
$\phi$ winds once round ${\cal M}$.  By continuity it follows that there must
be a point $p$ on the disk $D$ bounded by $C$ where $\phi = 0$.  By
translational symmetry
there is a line of points with $\phi = 0$.  This line is the center of the
cosmic string.  The cosmic string is a line of trapped potential energy.  In
order to minimize the total energy given the prescribed topology
(i.e.,
winding number), the thickness of the string (i.e., radius over which $V
(\phi)$
deviates significantly from 0) must be finite.  As first shown in Ref.
67, the
width $w$ of a string is
$$
w \simeq \lambda^{-1/2} \eta^{-1} \, , \eqno\eq
$$
from which it follows that the mass per unit length $\mu$ is
$$
\mu \simeq \eta^2 \, , \eqno\eq
$$
{i.e.,} independent of the self-coupling constant $\lambda$.
\par
Cosmic strings arise in any model in which the vacuum manifold satisfies the
topological criterion
$$
\Pi_1 ({\cal M}) \neq {\bf 1} \, . \eqno\eq
$$
Any field configuration $\phi
(\undertext{x})$ is characterized by an integer $n$, the element of $\Pi_1
({\cal M})$ corresponding to $\phi (\undertext{x})$. (Roughly speaking, $n$ can
be viewed
as the number of times the map $\varphi $ from {\cal C} to {\cal M} covers
{\cal M} ).
\par
A cosmic string is an example of a \undertext{topological defect}.  A
topological defect has a well-defined core, a region in space where $\phi
\not\in {\cal M}$ and hence $V (\phi) > 0$.  There is an associated
winding number, and it is quantized.  Hence, a topological defect is stable.
Furthermore, topological defects exist for theories with global and local
symmetry groups.

\section{Monopoles}
\par
If the
theory contains three real scalar fields $\phi_i$ with potential (3.46) (if
$|\phi|^2 = \sum\limits^3_{i=1} \, \phi^2_i$), then $\Pi_2 ({\cal M})
\neq
{\bf 1}$ and monopoles result.  The construction of a monopole
configuration is illustrated in Fig. 14.
As the origin in physical space we select a point which is to become
the center of the monopole.  Consider a sphere $S_r$ of radius $r$
surrounding this point.  A spherically symmetric monopole
configuration is obtained by the identity map
$$
\eqalign{
& S_r \rightarrow {\cal M} = S^2 \cr
& (r, \, \theta, \, \varphi) \, \mathop \rightarrow\limits_{\varphi}
\, (\theta, \, \varphi) \, .} \eqno\eq
$$
This configuration has winding number 1.  Since the winding number of
maps $S^2 \rightarrow S^2$ is quantized, it cannot change as $r$
varies.  Thus, the only way to obtain a single valued field
configuration at $r = 0$ is for $\varphi (r, \, \theta, \, \varphi)$
to leave ${\cal M}$ as $r \rightarrow 0$.  In particular, there is a
point ({e.g.,} $r = 0$) for which $\varphi = 0$.  This is the
center of the monopole.  We see that monopoles are topological
defects: they contain a core, have quantized winding number and are
stable.
\midinsert \vskip 8.5cm
\hsize=6in \raggedright
\noindent {\bf Figure 14.} Construction of a monopole: left is
physical space, right the vacuum manifold.  The field configuration
$\phi$ maps spheres in space onto ${\cal M}$.  However, a core region
of space near the origin is mapped onto field values not in ${\cal
M}$.
\endinsert

\section{Global Textures}

Next, consider a theory of four real scalar fields given by the Lagrangian
$$
{\cal L} = {1\over 2} \partial_\mu \phi \partial^\mu \phi - V (\phi) \eqno\eq
$$
with
$$
V (\phi) = {1\over 4} \lambda \, \left( \sum\limits^4_{i=1} \, \phi^2_i -
\eta^2 \right)^2 \, . \eqno\eq
$$
In this case, the vacuum manifold is ${\cal M} = S^3$ with topology
$$
\Pi_3 ({\cal M}) \neq {\bf 1} \, , \eqno\eq
$$
and the corresponding defects are the global textures$^{1,68,69)}$.
\midinsert \vskip 10cm
\hsize=6in \raggedbottom
\noindent{\bf Figure 15.} Construction of a global texture: left is
physical space, right the vacuum manifold. The field configuration
$\phi$ is a map from space to the vacuum manifold (see text).
\endinsert
\par
Textures, however, are quite different than the previous topological defects.
The texture construction will render this manifest (Fig. 15).  To construct a
radially symmetric texture, we give a field configuration $\phi (x)$ which
maps physical space onto ${\cal M}$.  The origin 0 in space (an arbitrary point
which will be the center of the texture) is mapped onto the north pole $N$ of
${\cal
M}$.  Spheres surrounding 0 are mapped onto spheres surrounding $N$.  In
particular, some sphere with radius $r_c (t)$ is mapped onto the equator
sphere of ${\cal M}$.  The distance $r_c (t)$ can be defined as the radius of
the texture.  Inside this sphere, $\phi (x)$ covers half the vacuum manifold.
Finally, the sphere at infinity is mapped onto the south pole of ${\cal M}$.
The configuration $\phi (\undertext{x})$ can be parameterized
by$^{69)}$
$$
\phi (x,y,z) = \left(\cos \chi (r), \> \sin \chi (r) {x\over r}, \>
\sin \chi (r) {y\over r}, \> \sin \chi (r) {z\over r} \right) \eqno\eq
$$
in terms of a function $\chi (r)$ with $\chi (0) = 0$ and $\chi (\infty) =
\pi$.  Note that at all points in space, $\phi (\undertext{x})$ lies in ${\cal
M}$.  There is no defect core.  All the energy is spatial gradient (and
possibly kinetic) energy.
\par
In a cosmological context, there is infinite energy available in an infinite
space.  Hence, it is not necessary that $\chi (r) \rightarrow \pi$ as $r
\rightarrow \infty$.  We can have
$$
\chi (r) \rightarrow \chi_{\rm max} < \pi \>\> {\rm as} \>\> r \rightarrow
\infty \, . \eqno\eq
$$
In this case, only a fraction
$$
n = {\chi_{\rm max}\over \pi} - {\sin 2 \chi_{\rm max}\over{2 \pi}} \eqno\eq
$$
of the vacuum manifold is covered:  the winding number $n$ is not quantized.
This is a reflection of the fact that whereas topologically nontrivial maps
from $S^3$ to $S^3$ exist, all maps from $R^3$ to $S^3$ can be deformed to
the trivial map.
\par
Textures in $R^3$ are unstable.  For the configuration described above, the
instability means that $r_c (t) \rightarrow 0$ as $t$ increases: the texture
collapses.  When $r_c (t)$ is microscopical, there will be sufficient energy
inside the core to cause $\phi (0)$ to leave ${\cal M}$, pass through 0 and
equilibrate at $\chi (0) = \pi$: the texture unwinds.
\par
A further difference compared to topological defects: textures are relevant
only for theories with global symmetry.  Since all the energy is in spatial
gradients, for a local theory the gauge fields can reorient themselves such as
to cancel the energy:
$$
D_\mu \phi = 0 \, . \eqno\eq
$$
\par
Therefore, it is reasonable to regard textures as an example of a new class of
defects, \undertext{semitopological defects}.  In contrast to topological
defects, there is no core, and $\phi (\undertext{x}) \epsilon {\cal M}$ for all
$\undertext{x}$.  In particular, there is no potential energy.  Second, the
winding number is not quantized, and hence the defects are unstable.  Finally,
they exist only in theories with a global internal symmetry.

\chapter{Formation and Evolution of Topological Defects}
\section{Kibble Mechanism}
\par
The Kibble mechanism$^{1)}$ ensures that in theories which admit
topological or semitopological defects, such defects will be produced
during a phase transition in the very early Universe.

Consider a mechanical toy model, first introduced by Mazenko, Unruh
and Wald$^{55)}$ in the context of inflationary Universe models, which
is useful in understanding the scalar field evolution.  Consider (see
Fig. 16) a lattice of points on a flat table.  At each point, a pencil
is pivoted.  It is free to rotate and oscillate.  The tips of nearest
neighbor pencils are connected with springs (to mimic the spatial
gradient terms in the scalar field Lagrangian).  Newtonian gravity
creates a potential energy $V(\varphi)$ for each pencil ($\varphi$ is
the angle relative to the vertical direction).  $V(\varphi)$ is
minimized for $| \varphi | = \eta$ (in our toy model $\eta = \pi /
2$).  Hence, the Lagrangian of this pencil model is analogous to that
of a scalar field with symmetry breaking potential (3.46).

\midinsert \vskip 5.5cm
\hsize=6in \raggedbottom
\noindent{\bf Figure 16.} The pencil model: the potential energy of a
simple pencil has the same form as that of scalar fields used for
spontaneous symmetry breaking.  The springs connecting nearest
neighbor pencils give rise to contributions to the energy which mimic
spatial gradient terms in field theory.
\endinsert

At high temperatures $T \gg T_c$, all pencils undergo large amplitude
high frequency oscillations.  However, by causality, the phases of
oscillation of pencils with large separation $s$ are uncorrelated.
For a system in thermal equilibrium, the length $s$ beyond which
phases are random is the correlation length $\xi (t)$. By causality there
is an apriori causality bound on
$\xi$:
$$
\xi (t_c) < t_c \, , \eqno\eq
$$
where $t_c$ is the causal horizon, at temperature $T_c$.

The critical temperature $T_c$ is the temperature at which the
thermal energy is equal to the energy a pencil needs to jump from
horizontal to vertical position.  For $T < T_c$, all pencils want to
lie flat on the table.  However, their orientations are random beyond
a distance of $\xi (t_c)$.

The boundaries between the domains of correlated orientation become
topological defect. Hence, it follows from the above causality argument
that during the phase transition a network of defects with mean separation
$\xi (t) \leq t$ will form.

For models of structure formation and for defects formed in grand unified
phase transitions we are interested in models with a scale of symmetry
breaking $\eta \sim 10^{16} GeV$ corresponding to a time of formation
$t_{c} \sim 10^{-35} sec$.

The evolution of the field configuration (and thus of the network of
defects) may be very complicated. However, the same causality argument as
used above tells us that no correlations on scales $>t$ can be
established (see, however, the caveats of Refs. 70 and 71). Hence, even at
times $t \gg t_{c}$, a network of defects will persist with
$$\xi (t) \leq t .\eqno\eq $$
For some applications, it is important to have a realistic estimate for the
initial separation of defects rather than the general causality bound
(4.1). In order to improve on (4.1) we must assume that $\varphi$ is in
thermal equilibrium. Our goal is to estimate $\xi (t)$ once the defects
have stabilized after the phase transition.

We follow the methods of Refs. 1,3 and 72. The first step of the analysis
is to calculate the location $\varphi_{m}(T)$ of the minimum of the
finite temperature effective potential $V_{T}(\varphi)$. From (3.16) it
follows that for $T \leq T_{c}$
$$\varphi_m(T) = \eta \left(1-\left( {T\over T_c}\right) ^2\right) ^{1/2}.
\eqno\eq $$
The difference in free energy between $\varphi = \varphi_{m}(T)$ and
$\varphi = 0$ is
$$\Delta V(T) = V(0) -V_T(\varphi_m(T))={1\over 4}\lambda\eta^4\left(1
-\left( {T\over T_c}\right) ^2\right) ^2. \eqno\eq $$
The proto-domain size is given by equating spatial gradient and potential
energy density
$$\left( {\varphi _m(T)\over\xi }\right) ^2\simeq\lambda\eta^4\left(
1-\left( {T\over T_c}\right)^2\right)^2 \eqno\eq $$
giving
$$\xi (T)\simeq\lambda^{-1/2}\eta^{-1}\left( 1-\left( {T\over
T_c}\right)^2\right)^{-1/2} .\eqno\eq $$
The Ginsburg temperature $T_{G}$ is defined as the temperature below which
a thermal fluctuation has insufficient energy to take a correlation
volume of `true vacuum' $\varphi=\varphi_{m}(T)$ into the symmetric
configuration $\varphi = 0$. Thus, the criterion for $T_{G}$ is
$$\xi^3(T_G)\Delta V(T_G)=T_G . \eqno\eq $$
Inserting (4.4) and (4.6) we obtain
$$ T_G = \lambda^{-1/2}\eta\left( 1-\left( {T_G\over T_c
}\right)^2\right)^{1/2}\eqno\eq $$
and therefore (from (4.6))
$$\xi (T_G)=\lambda^{-1}T_G^{-1}\sim\lambda^{-1}\eta^{-1} . \eqno\eq $$
As a result of the above analysis, it follows that the network of
topological defects `freezes out' (i.e. is not continuously destroyed and
recreated by thermal fluctuations) at time $t_{G}$  corresponding to the
temperature $T_{G}$. The correlation length at the Ginsburg temperature is
microscopic, i.e. proportional to $\eta^{-1}$, and thus much smaller than
the naive causality argument (4.1) would indicate. This fact will be
important for cosmic string driven baryogenesis.

Note that the Kibble mechanism was discussed above in the context of a
global symmetry breaking scenario.  As pointed out in Ref. 73, the
situation is more complicated in local theories in which gauge field
can cancel spatial gradients in $\varphi$ in the energy functional,
and in which spatial gradients in $\varphi$ can be gauged away.
Nevertheless, as demonstrated numerically (in $2 + 1$ dimensions) in
Ref. 74 and shown analytically in Ref. 75, the Kibble mechanism also
applies to local symmetries.

\section {Domain Wall and Monopole Problems}\par
\par
As stated in Table 1, models with domain walls and local monopoles are
ruled out on cosmological grounds. In both cases, the problem is that the
energy density in the defects dominates over the energy density in
radiation already at the time of nucleosynthesis; in other words, the
defects would overclose the Universe. The reasons why this problem occurs,
however, are different for domain walls and local monopoles.

Let us demonstrate explicitly why stable domain walls are a
cosmological disaster$^{76)}$.  If domain walls form during a phase transition
in the early Universe, it follows by causality (see however the caveats
of Refs. 70 and 71) that even today there will be at least one wall
per Hubble volume.  Assuming one wall per Hubble volume, the energy
density $\rho_{DW}$ of matter in domain walls is
$$
\rho_{DW} (t) \sim \eta^3 t^{-1} \, , \eqno\eq
$$
whereas the critical density $\rho_c$ is
$$
\rho_c = H^2 \, {3\over{8 \pi G}} \sim m^2_{p\ell} \, t^{-2} \, .
\eqno\eq
$$
Hence, for $\eta \sim 10^{16}$ GeV the ratio of (4.10) and (4.11) is
$$
{\rho_{DW}\over \rho_c} \, (t) \sim \, \left({\eta\over{m_{p\ell}}}
\right)^2 \, (\eta t) \sim 10^{52} \, . \eqno\eq
$$

The above argument depends in an essential way on the dimension of the
defect.  One cosmic string per Hubble volume leads to an energy
density $\rho_{cs}$ in string
$$
\rho_{cs} \sim \eta^2 \, t^{-2} \, , \eqno\eq
$$
which scales like the background radiation density.
Later in this chapter we shall see that the scaling (4.13) does indeed hold in
the
cosmic string model.  Hence, cosmic strings do not lead to
cosmological problems.  In the contrary, since for GUT models with
$\eta \sim 10^{16}$ GeV
$$
{\rho_{cs}\over \rho_c} \sim \, \left({\eta\over m_{p \ell}} \right)^2
\sim 10^{-6} \, , \eqno\eq
$$
cosmic strings in these models could provide the seed perturbations
responsible for structure formation.

Theories with local monopoles are ruled out on cosmological
grounds$^{77)}$ (see again the caveats of Refs. 70 and 71) for
rather different reasons.  Since there are no long range forces
between local monopoles, their number density in comoving coordinates
does not decrease.  Since their contribution to the energy density
scales as $a^{-3} (t)$, they will come to dominate the mass of the
Universe, provided $\eta$ is sufficiently large.

Theories with global monopoles$^{78)}$ are not ruled out, since
there are long range forces between monopoles which lead to a
``scaling solution" with a fixed number of monopoles per Hubble
volume.

\section{Cosmic String Evolution}
\par
Applied to cosmic strings, the Kibble mechanism implies that at the
time of the phase transition, a network of cosmic strings with typical
step length $\xi (t_G)$ will form.  According to numerical
simulations$^{79)}$, about 80\% of the initial energy is in infinite
strings and 20\% in closed loops.

The evolution of the cosmic string network for $t > t_G$ is
complicated .  The key processes are loop production
by intersections of infinite strings (see Fig. 17) and loop shrinking
by gravitational radiation.  These two processes combine to create a
mechanism by which the infinite string network loses energy (and
length as measured in comoving coordinates).  It will be shown that as
a consequence, the correlation length of the string network is always
proportional to its causality limit
$$
\xi (t) \sim t \, . \eqno\eq
$$
Hence, the energy density $\rho_\infty (t)$ in long strings is a fixed
fraction of the background energy density $\rho_c (t)$
$$
\rho_\infty (t) \sim \mu \xi (t)^{-2} \sim \mu t^{-2} \eqno\eq
$$
or
$$
{\rho_\infty (t)\over{\rho_c (t)}} \sim G \mu \, . \eqno\eq
$$
\midinsert \vskip 3.5cm
\hsize=6in \raggedright
\noindent {\bf Figure 17.}  Formation of loops by self intersection of
infinite strings.  According to the original cosmic string scenario, loops
form with radius $R$ determined by the instantaneous correlation length of the
infinite string network.
\endinsert
\par
We conclude that the cosmic string network approaches a ``scaling
solution"$^{80)}$ in which the statistical properties of the
network are time independent if all distances are scaled to the
horizon distance.

The origin of the scaling solution for the infinite string network can be
understood heuristically as follows. If the curvature radius $\xi (t)$ of
this network is much larger than the Hubble radius $t$, the network
will be frozen in comoving coordinates (since the Hubble damping term
dominates in the equations of motion). Hence in the radiation dominated
FRW epoch
$$\xi (t) \sim a(t)\sim t^{1/2} \eqno\eq $$
and the Hubble radius will catch up to $\xi (t)$. Conversely, if $\xi (t)
\ll t$ then the tension term in the equations of motion for the string
will dominate, the strings will oscillate relativistically and there will
be frequent self intersections of the strings, leading to rapid loop
production and to increasing $\xi(t)/ t$. Combining these two arguments, we
conclude that there must be a `dynamical fixed point' with $\xi (t)\sim
t$.

A first step in a more rigorous analysis of cosmic string evolution is the
derivation of the effective equation of motion for the strings, Note that
this equation must follow from the field equations since the string is
merely a particular topologically stable field configuration.

The equations of motion of a string can be derived from the Nambu action
$$
S = - \mu \int d \sigma d \tau \left(- \det g^{(2)}_{ab} \right)^{1/2} \> \>
a, b = 0, 1 ,\eqno\eq
$$
where $g^{(2)}_{ab}$ is the world sheet metric and $\sigma$ and
$\tau$ are the world sheet coordinates.  In flat space-time, $\tau$
can be taken to be coordinate time, and $\sigma$ is an affine
parameter along the string.  In terms of the string
coordinates $X^\mu (\sigma, \tau)$ and the metric $g^{(4)}_{\mu\nu}$ of
the background space-time,
$$
g^{(2)}_{ab} = X^\mu_{,a} X^\nu_{,b} g^{(4)}_{\mu\nu} \, . \eqno\eq
$$
{}From
general symmetry considerations,  it is possible to argue that the
Nambu action is the correct action.  However, I shall follow
Foerster$^{81)}$ and
Turok$^{82)}$ and give a direct heuristic derivation.  We start from a
general quantum field theory Lagrangian ${\cal L}_{QFT}$.  The action is
$$
S = \int d^4 y {\cal L}_{QFT} \left(\phi (y)\right)\eqno\eq
$$
We assume the existence of a linear topological defect at $X^\mu (\sigma,
\tau)$.  The idea is to change variables so that $\sigma$ and $\tau$ are
two of the new coordinates, and to expand $S$ to lowest order in $w/R$, where
$w$ is the width of the string and $R$ its curvature radius.  As the other new
coordinates we take coordinates $\rho^2$ and $\rho^3$ in the normal plane to
$X^\mu (\sigma, \tau)$.  Thus the coordinate transformation takes the
coordinates $y^\mu (\mu = 0, 1, 2, 3)$ to new ones $\sigma^\nu = (\tau, \sigma,
\rho^2, \rho^a)$:
$$
y^\mu (\sigma^a) = X^\mu (\sigma, \tau) + \rho^i n^\mu_i (\sigma,
\tau)\eqno\eq
$$
where $i = 2,3$ and $n^\mu_i$ are the basis vectors in the normal plane to the
string world sheet.  The measure transforms as
$$
\int d^4 y = \int d \sigma d \tau d \rho^2 d \rho^3 (\det M_a^\mu)\eqno\eq
$$
with
$$
M^\mu_a = \, {\partial y^\mu\over{\partial \sigma^a}} = \, \pmatrix{\partial
X^\mu/\partial (\sigma, \tau)\cr
n^\mu_i\cr} + O (\rho)\, .\eqno\eq
$$
The determinant can easily be evaluated using the following trick
$$
\det M^\mu_a = \, \left( - \det \eta_{\mu \nu} M^\nu_a M^\nu_b \right)^{1/2}
\equiv \sqrt{- \det D_{ab}}\eqno\eq
$$
$$
D = \, \pmatrix{{\partial x^\mu\over{\partial (\sigma, \tau)}} \, {\partial
X^\nu\over{\partial (\sigma, \tau)}} \eta_{\mu \nu} & {\partial X^\mu\over
{\partial (\sigma , \tau)}} n^\nu_b \eta_{\mu\nu}\cr
{\partial X^\mu\over{\partial (\sigma, \tau)}} n^\nu_a \eta_{\mu\nu} & n^\mu_a
n^\nu_b \eta_{\mu\nu}\cr} = \pmatrix{X^\mu_{,a} X^\nu_{,b} \eta_{\mu\nu} & 0\cr
0 & \delta_{ab}\cr} + 0 \, \left({w\over R}\right) \eqno\eq
$$
Hence
$$
\eqalign{S &= \int d \sigma d \tau \left( - \det g^{(2)}_{ab} \right)^{1/2}
\int d \rho^2 d \rho^3 {\cal L} (y (\sigma, \tau, \rho^2, \rho^3)) + O
\left({w\over R} \right)\cr
&= - \mu \int d \sigma d \tau \left( - \det g^{(2)}_{ab} \right)^{1/2} + O
\left({w\over R}\right)\, . }\eqno\eq
$$
Here, $- \mu$ is the integral of {\cal L} in the normal plane of $X$.  To first
order in $w/R$, it equals the integral of $-{\cal H}$; hence it is the mass per
unit length.
\par
This derivation of the Nambu action is instructive as it indicates a method
for calculating corrections to the equations of motion of the string when
extra fields are present, \eg\ for superconducting cosmic strings.  It also
gives a way of calculating the finite thickness corrections to the equations
of motion which will be important at cusps (see below).
\par
In flat space-time we can consistently choose $\tau = t, \dot x \cdot x^\prime
= 0$ and $\dot x^2 + x^{\prime^2} = 0$.  The equations of motion derived from
the Nambu action then become
$$
\ddot {\undertext{x}} - \undertext{x}^{\prime\prime} = 0\, . \eqno\eq
$$
where $\prime$ indicates the derivative with respect to $\sigma$. As expected,
we obtain
the relativistic wave equation.  The general
solution can be decomposed into a left moving and a right moving mode
$$
\undertext{x} (t, \sigma) = {1\over 2} \, \left[ \undertext{a} (\sigma - t) +
\undertext{b} (\sigma + t ) \right] \eqno\eq
$$
The gauge conditions imply
$$
\dot {\undertext{a}}^2 = \dot {\undertext{b}}^2 = 1\eqno\eq
$$
For a loop, $\undertext{x} (\sigma, t)$ is periodic and hence the time average
of $\dot {\undertext{a}}$ and $\dot {\undertext{b}}$ vanish. Thus, $\dot
{\undertext{a}}$
and $\dot {\undertext{b}}$ are closed curves on the unit sphere with vanishing
average.  Two such curves generically intersect if they are
continuous$^{83)}$.  An intersection corresponds to a point with
$\undertext{x}^\prime = 0$ and $\dot {\undertext{x}} = 1$.  Such a point moving
at the speed of light is called a cusp. Note that $\dot {\undertext{x}}
(\sigma, t)$
need
not be continuous.  Points of discontinuity are called kinks.  Both
cusps and kinks will be smoothed out by finite thickness
effects$^{84)}$.
\par
The Nambu action does not describe what happens when two strings hit.  This
process has been studied numerically for both global$^{85)}$ and
local$^{86)}$ strings.  The authors of those papers set up scalar field
configurations
corresponding to two strings approaching one another and evolve the complete
classical scalar field equations.  The result of the analysis is that strings
do not cross but exchange ends, provided the relative velocity is smaller than
0.9.  Thus, by self intersecting, an infinite string will split off a loop
(Figure 17).  An important open problem is to understand this process
analytically.  For a special value of the coupling constant Ruback has given a
mathematical explanation$^{87)}$ (see also Shellard and
Ruback in Ref. 86).
\par
There are two parts to the nontrivial evolution of the cosmic string network.
Firstly, loops are produced by self intersections of infinite strings.  Loops
oscillate due to the tension and slowly decay by emitting gravitational
radiation$^{88)}$.  Combining the two steps we have a process by which energy
is transferred from the cosmic string network to radiation.
\par
There are analytical indications that a stable ``scaling solution"
for the
cosmic string network exists$^{4)}$.  It is given by on the order 1 infinite
string segment crossing every Hubble volume.  The correlation length $\xi (t)$
of an infinite string is of the order $t$.  Hence, at time $t$ loops of radius
$R \sim t$ are produced, of the order 1 loop per Hubble volume per expansion
time.  A heuristic argument for the scaling solution is due to Vilenkin.  Take
$\tilde \nu (t)$ to be the mean number of infinite string segments per Hubble
volume.  Then the energy density in infinite strings is
$$
\rho_\infty (t) = \mu \tilde \nu (t) t^{-2} \eqno\eq
$$
The number of loops $n(t)$ produced per unit volume is given by
$$
{d n (t)\over{dt}} = c \tilde \nu^2 t^{-4} \eqno\eq
$$
where $c$ is a constant of the order $1$.  Conservation of energy in strings
gives
$$
{d \rho_\infty (t)\over{dt}} + {3\over{2 t}} \, \rho_\infty (t) = - c^\prime
\mu t \, {dn\over{dt}} = - c^\prime \mu \tilde \nu^2 t^{-3} \eqno\eq
$$
or, written as an equation for $\tilde \nu (t)$
$$
\tilde {\dot \nu} - \, {\tilde \nu\over{2 t}} = - cc^\prime \tilde \nu^2 t^{-
1}\eqno\eq
$$
Thus if $\tilde \nu \gg 1$ then $\tilde {\dot \nu} < 0$ while if $\tilde \nu
\ll 1$ then $\tilde {\dot \nu} > 0$.  Hence there will be a stable solution
with $\tilde \nu \sim 1$.

The precise value of $\tilde \nu$ must be determined in numerical simulations.
These simulations are rather difficult because of the large dynamic range
required and due to singularities which arise in the evolution equations near
cusps.  In the radiation dominated epoch, $\tilde \nu$ is still uncertain by a
factor of about 10.  The first results were reported in Ref. 89.
More recent results are due three groups.
Bennett and Bouchet$^{90)}$ and Allen and Shellard$^{91)}$ are
converging on
a value $10 < \tilde \nu < 20$, whereas Albrecht and Turok$^{92)}$ obtain a
value which is about 100.

\section{Scaling Solution for Strings}
\par
The scaling solution for the infinite strings implies that the network of
strings looks the same at all times when scaled to the Hubble radius.  This
should also imply that the distribution of cosmic string loops is scale
invariant in the same sense.  At present, however, there is no convincing
evidence from numerical simulations that this is really the case.
\par
A scaling solution for loops implies that the distribution of $R_i (t)$, the
radius of loops at the time of formation, is time independent of time after
dividing
by $t$.  To simplify the discussion, I shall assume that the
distribution in monochromatic, \ie\
$$
R_i (t)/t = \alpha\, . \eqno\eq
$$
{}From Figure 17, we expect $\alpha \sim 1$.  The numerical simulations,
however, now give $\alpha < 10^{-2}$~$^{90,91)}$.
\par
{}From the scaling solution (4.15) for the infinite strings we can derive the
scaling solution for loops.  We assume that the energy density in long strings
-- inasmuch as it is not redshifted -- must go into loops.  $\beta$ shall be a
measure for the mean length $\ell$ in a loop of ``radius" $R$
$$
\ell = \beta R\, . \eqno\eq
$$
Since per expansion time and Hubble volume about 1 loop of radius $R_i (t)$ is
produced, we know that the number density in physical coordinates of loops of
radius $R_i (t)$ is
$$
n (R_i (t), t) = ct^{-4}\eqno\eq
$$
with a constant $c$ which can be calculated from (4.31), (4.35) and
(4.36).
Neglecting gravitational radiation, this number density simply redshifts
$$
n (R,t) = \, \left({z (t)\over{z (t_f (R))}} \right)^3 n (R, t_f (R))\, ,
\eqno\eq
$$
where $t_f (R)$ is the time when loops of radius $R$ are formed.  Isolating
the $R$ dependence, we obtain
$$
n (R, t) \sim R^{-4} z (R)^{-3}\eqno\eq
$$
where $z (R)$ is the redshift at time $t=R$.  We have the following special
cases:
$$
\eqalign{n (R, t) \sim R^{-5/2} t^{-3/2} \qquad & t < t_{eq}\cr
n (R, t) \sim R^{-5/2} t_{eq}^{1/2} t^{-2} \qquad & t > t_{eq} \, , \, t_f
(R) < t_{eq}\cr
n (R, t) \sim R^{-2} t^{-2} \qquad & t > t_{eq} \, , \, t_f (R) > t_{eq} \,
.}\eqno\eq
$$
\par
The proportionality constant $c$ is
$$
c = {1\over 2} \beta^{-1} \alpha^{-2} \tilde \nu\eqno\eq
$$
(see \eg\ Ref. 93).  In deriving (4.41) it is important to note that $n (R_i
(t), t) dR_i$ is the number density of loops in the radius interval $[R_i ,
R_i + dR_i]$.  Hence, in the radiation dominated epoch
$$
n (R, t) = \nu R^{- 5/2} t^{-3/2} \eqno\eq
$$
with
$$
\nu = {1\over 2} \beta^{-1} \alpha^{1/2} \tilde \nu \, .\eqno\eq
$$
\par
{}From (4.43) we can read off the uncertainties in $\nu$ based on the
uncertainties in the numerical results.  Both $\alpha^{1/2}$ and $\tilde \nu$
are determined only up to one order of magnitude.  Hence, any quantitative
results which depend on the exact value of $\nu$ must be treated with a grain
of salt.
\par
Gravitational radiation leads to a lower cutoff in $n (R, t)$.  Loops with
radius smaller than this cutoff were all formed at essentially the same time
and hence have the same number density.  Thus, $n (R)$ becomes flat.  The
power in gravitational radiation $P_G$ can be estimated using the quadrupole
formula$^{94)}$.  For a loop of radius $R$ and mass $M$
$$
P_G = {1\over 5} G < \dot{\ddot Q} \dot{\ddot Q} > \, , \eqno\eq
$$
where $Q$ is the quadrupole moment, $Q \sim MR^2$, and since the frequency of
oscillation is $\omega = R^{-1}$
$$
P_G \sim G (MR^2)^2 \omega^6 \sim (G \mu) \mu \, . \eqno\eq
$$
\par
Even though the quadrupole approximation breaks down since the loops move
relativistically, (4.45) gives a good order of magnitude of the power of
gravitational radiation.  Improved calculations$^{88)}$ give
$$
P_G = \gamma (G \mu) \mu\eqno\eq
$$
with $\gamma \sim 50$.  (4.36) and (4.46) imply that
$$
\dot R = \tilde \gamma G \mu\eqno\eq
$$
with $\tilde \gamma \equiv \gamma/\beta \sim 5$ (using $\beta \simeq
10$~$^{89)}$).  Note that the rate of decrease is constant.  Hence,
$$
R (t) = R_i - (t - t_i) \tilde \gamma G \mu\eqno\eq
$$
and the cutoff loop radius is
$$
R_c \sim \tilde \gamma G \mu t_i\, . \eqno\eq
$$
\par
Let us briefly summarize the scaling solution:
\item{1)} At all times the network of infinite strings looks the same when
scaled by the Hubble radius.  A small number of infinite string segments cross
each Hubble volume and $\rho_\infty (t)$ is given by (4.31).
\item{2)}  There is a distribution of loops of all sizes $0 \le R < t$.
Assuming scaling for loops, then
$$
n (R, t) = \nu R^{-4} \, \left({z (t)\over{z (R)}}\right)^3 \, , \> R
\> \epsilon \> [ \tilde \gamma G \mu t,  \alpha t]\eqno\eq
$$
where $\alpha^{-1} R$ is the time of formation of a loop of radius $R$.  Also
$$
n (R, t) = n ( \tilde \gamma G \mu t,  t) \, , \> R < \tilde \gamma G \mu t\,
. \eqno\eq
$$

Although the qualitative characteristics of the cosmic string scaling
solution are well established, the quantitative details are not.  The
main reason for this is the fact that the Nambu action breaks down at
kinks and cusps.  However, kinks and cusps inevitably form and are
responsible for the small scale structure on strings.  In fact, coarse
graining by integrating out the small scale structure may give an
equation of state for strings which deviates from that of a Nambu
string$^{95)}$.  Attempts at understanding the small scale structure
on strings are at present under way$^{96)}$.

\section{Scaling Solution for Textures}
\par
Applied to textures $^{69)}$, the Kibble mechanism implies that on all scales
$r \geq \xi(t_{G})$, field configurations with nonvanishing winding number
$n_{w}$ are frozen in after the phase transition at time $t_{G}$.

The general causality argument implies that at all times $t > t_{G}$, on
scales greater than $t$ the field configuration will be random, and hence
configurations with winding number $n_{w} \neq 0$ will persist.

The texture dynamics is determined by the field equation
$$\ddot{\varphi } +3H\dot{\varphi } -a^{-2} \nabla ^2 \varphi = -V'(\varphi ).
 \eqno\eq $$
The second term on the left hand side is proportional to $H^{2} \varphi$,
the third term scales as $k^{2}a^{-2} \varphi$, where $k^{-1}$ is the
wavelength of the inhomogeneity of the field configuration. Comparing
these two terms, we see that for
$$ ka^{-1} < H , \eqno\eq $$
i.e. on scale larger than the Hubble radius, a texture configuration will
be frozen in, whereas for
$$ka^{-1} > H ,\eqno\eq $$
on scales smaller that the Hubble radius, the motion of $\varphi$ will be
relativistic, and the texture configuration will become homogeneous on a
time scale of $t$.

Hence, in the texture scenario, at time $t\gg t_{G}$, the field
configuration $\varphi (\underline{x})$ is predicted to be homogeneous on
scales smaller than $t$, but inhomogeneous on larger scales. There is a
finite (and time independent) probability $p(n_{c})$ that per horizon
volume there is a field configuration with winding $n_{w}$ greater than
$n_{c}$. The probability $p(n_{c})$ can be determined using combinatorial
arguments $^{97)}$.

The evolution of texture configurations has been studied numerically in
Refs. 98-100 and analytically in Ref. 101 (see also Ref. 102). The
qualitative features of the evolution depend on wether $n_{w}$ is greater
or smaller than the critical winding $n_{c}$. If $n_{w}> n_{c}$, then the
texture will contract, the winding will increase, and eventually the texture
will unwind by $\varphi$ jumping over the potential
barrier at the texture center. On the other hand, if $n_{w}< n_{c}$, the
field configuration will dissipate. It will expand with $n_{w}$ decreasing
to zero continuously.

The physics of texture evolution can be most readily analyzed by studying
the spherically symmetric configuration (3.55) with $\chi$ a function of
$r$ and $t$. For such a configuration, the action in an expanding FRW
background metric is
$${\cal S} = \int d^4x\; r^2a^3[\dot{\chi }^2 -a^{-2}{\chi '}^2
-2a^{-2}r^{-2}\sin^2\chi ].
 \eqno\eq $$
The second term corresponds to the radial gradient energy, the third to
angular gradient energy.

Consider a field configuration as sketched in Fig. 18 a with $\chi_{max} >
n_{c} > \pi/2$. In this case, the forces $F_{1}$ and $F_{2}$ act as
indicated in the figure. In order to minimize the angular gradient
energy, $\chi_{max}$ wants to increase (see F1). However, for $r < x$
there is an  additional (and in part opposing) force $F_{2}$: in order to
reduce the angular gradient energy, $r(\chi)$ will tend to decrease, i.e.
the field configuration will contract and its total winding will
increase.

However, for $\chi_{max} < \pi/2$, the angular gradient force tends to
reduce $\chi_{max}$, and to reduce the radial gradients, $\chi(r)$ will
tend to decrease even for $\chi < \chi_{max}$. Hence, the field
configuration will dissipate (see Figure 18 b).

If $n_{w}$ is only slightly larger than $\pi/2$, the force $F1$ of Fig.15
a is too weak to offset the force F2 described above. The critical winding
$n_{c}$ is hence larger than $0.5$. The precise value has been determined
in Refs. 99 and 101. It depends on the length of the `plateau region' of
$\chi (r)$ (which is the inter-texture separation) and on the expansion
rate of the Universe. For realistic parameters $^{101)}$
$$0.65 < n_c < 0.75\;\; .\eqno\eq $$
As will be shown in a later Chapter, only textures with $n_{w} > n_{c}$
generate localized density perturbations which can act as seeds for cosmic
structure formation.
\goodbreak \midinsert \vskip 4.5cm
\hsize=6in \raggedright
\noindent {\bf Figure 18.}
A sketch of the forces acting on a radially symmetric texture configuration and
which cause unwinding in Fig. 18a if  $n_w>n_c$, and dissipation if $n_w<n_c$
(Fig. 18b).
\endinsert
\par
\chapter{Introduction to Structure Formation}
\section{Power Spectrum}
\par
In Chapter 2, the cosmology of a homogeneous and isotropic Universe was
reviewed. In order to understand structure formation, it is essential to
study the evolution of inhomogeneities at a linearized level. This will be
adequate to understand the early evolution of density perturbations in the
Universe.

Starting point of the relativistic theory of cosmological perturbations
$^{103)}$ are the linearized Einstein equations. If we take the general
Einstein equations
$$G_{\mu\nu }=8\pi G \; T_{\mu\nu } , \eqno \eq $$
where $G_{\mu v} (g_{\alpha \beta})$ is the Einstein tensor and $T_{\mu
v}$ is the energy-momentum tensor of matter, and expand about a
cosmological background solution
$$ {g_{\mu\nu}}^{(0)}=diag\left(1,-a^2(t),-a^2(t),-a^2(t)\right) \eqno\eq $$
and
$${T^{\mu}_{\nu}}^{(0)}=diag\left( \rho,-p,-p,-p\right) \eqno\eq $$
(see Chapter 2), then we obtain the linearized equations
$$\delta G_{\mu\nu}\left( {g_{\alpha\beta}}^{(0)}+h_{\alpha\beta}\right)=8\pi
G\;
\delta T_{\mu\nu} . \eqno\eq $$
Equation (5.4) relates the perturbation $h_{\mu v}$ of the metric,
i.e.
$$h_{\mu\nu}=g_{\mu\nu}- {g_{\mu\nu}}^{(0)}, \eqno\eq $$
to the matter perturbations.

To gain a heuristic understanding of how the perturbations evolve, recall
that gravity is a purely attractive force. Given an initial mass
perturbation $\delta m$, the force on surrounding particles will be
$$F\sim\delta m\eqno\eq .$$
Since (neglecting for a moment the expansion of the Universe)
$$\delta\ddot{m}\sim F,\eqno\eq $$
we see that in a nonexpanding background the growth of perturbations is
exponential. In an expanding background, there will be a damping term
depending on $H$. Hence, perturbations will increase only as a power of
time.

The details of the analysis are rather complicated (see Refs. 30 and 104
for recent reviews). The result is that the density contrast $\delta \rho$
grows as follows
$$\delta\rho (t)\sim\cases{
t^{2/3} & $t>t_{eq}$ \cr
t & $t<t_{eq}$ , $\lambda >t$ \cr
{\rm const}& $t<t_{eq}$ , $\lambda <t$}  \eqno\eq$$
Note that on length scales $\lambda$ greater than the Hubble radius $t$,
the quantity $\delta \rho$ is not gauge invariant, i.e. it depends on the
slicing of space-time $^{104)}$. The quantity which is gauge invariant is the
relativistic potential $\Phi$, which is time independent if the equation
of state of the background cosmology is constant $^{104)}$. In a gauge in which
$g_{\mu v}$ is diagonal and for models of matter in which $\delta T_{i j}$
is diagonal at linearized level (a condition satisfied by most
interesting models of matter), $\Phi$ can be identified as follows:
$$g_{\mu\nu}=(1+2\Phi )dt^2-a^2(t)(1-2\Phi )d\underline{x}^2 . \eqno\eq $$
We will use the results of (5.8) when describing the evolution of the
power spectrum.

The main quantity
of interest is the r.m.s. mass excess $(\delta M/M) \, (k, t)$ at time $t$ in
a sphere of radius $k^{-1}$.  Given a smooth density distribution
$$
\rho ( \undertext{x} , t) = \rho_0 (t) + \delta \rho ( \undertext{x} , t) \, ,
\eqno\eq
$$
the r.m.s. mass excess can be related to the Fourier mode $\delta \rho (k)$ in
a straightforward manner$^{18)}$.  The result is
$$
\left( {\delta M\over M} \right)^2 \, (k, t) \simeq k^3 \big| {\delta
\rho\over \rho_0} \big|^2 \, (k,t) \, . \eqno\eq
$$
The adopted convention for Fourier transformation is
$$
\delta \rho (\undertext{x}) = (2 \pi)^{- 3/2} V^{-1/2} \int d^3 k e^{i
\undertext{k} \undertext{x}} \delta \rho (\undertext{k}) \, . \eqno\eq
$$
The result (5.11) holds provided $|\delta\rho (k)|^2$ is proportional to $k^n$
with
$n > -3$.  An intuitive way to understand the result is as follows:
perturbations with wave number larger than $k$ average to zero in a volume
$k^{-3}$, perturbations with wave number smaller then $k$ are phase space
suppressed such that $(\delta M / M) \, (k)$ receives its major contribution
from Fourier modes of wave number $k$.  Their phase space volume is $k^3$.

The most commonly used function describing the ensemble of
perturbations is the spectrum $P(k)$.  By definition, the power
spectrum is the square of the modulus of the Fourier space density
contrast
$$
P (k) = \big| {{\delta \rho} \over \rho_0} (k) \big|^2 \, . \eqno\eq
$$
Hence (from (5.11)), $P(k)$ is related to the r.m.s. mass fluctuations
$(\delta M/M)\, (k,t)$ on physical length scale
$$
\lambda_k = a (t) \, {2\pi\over k} \eqno\eq
$$
at time $t$ via
$$
\left( {\delta M\over M} \right)^2 \, (k,t) \simeq k^3 P (k) \, .
\eqno\eq
$$

The scaling solution for topological defect models implies that when
measured at the time $t_H (k)$ when the wavelength $\lambda_k$ equals
the Hubble radius, the r.m.s. mass perturbation $(\delta M/M) \, (k)$
is independent of $k$, i.e.,
$$
{\delta M\over M} \, (k, t_H (k)) = {\rm const} \, . \eqno\eq
$$
This is because at any time $t$, a constant fraction of the mass $M$
inside the Hubble radius is contained in the topological defects.  For
example, one cosmic string of length $t$ contains mass $\delta M= \mu
t$ compared to the total mass $M \sim t^3 \rho (t) \sim t$ inside the
Hubble radius, thus leaving the ratio $\delta M/M$ time independent.

Eq. (5.16) is the same result as is obtained for inflationary Universe
models.  Hence, we conclude that all three main models of structure
formation: adiabatic random phase perturbations from inflation, cosmic
strings, and global textures, to a first approximation produce a
scale-invariant spectrum.

To convert (5.16) into an expression for the power spectrum $P(k)$, we
use the fact that $\delta M/M$ grows as the scale factor $a(t)$ during
the matter dominated epoch on scales smaller than the Hubble
radius (see (5.8))
$$
{\delta M\over M} \, (k,t) = \left({t\over{t_H (k)}} \right)^{2/3} \,
{\delta M\over M} \, (k, t_H (k)) \, . \eqno\eq
$$
On scales larger than the Hubble radius at $t_{eq}$
$$
t_H (k) = 2 \pi k^{-1} a (t_H (k)) \sim t_H^{2/3} (k) \, k^{-1} \, ,
\eqno\eq
$$
and hence
$$
t_H (k) \sim k^{-3} \, . \eqno\eq
$$
Therefore, combining (5.16), (5.17) and (5.19)
$$
{\delta M\over M} \, (k,t) \sim k^2 \, . \eqno\eq
$$
{}From (5.15) it follows that
$$
P (k) \sim k^n \eqno\eq
$$
with $n = 1$.

Recently, there has been some interest in deviations from scale
invariance.  In models of inflation, a deviation comes about$^{105)}$
because $H$ decreases slowly during inflation.  In topological defect
models, numerical$^{106,107)}$ and semi-analytical$^{108)}$ studies
have also shown small deviations from scale invariance.  These
deviations, however, all small and quite model dependent.

\section{CMB Anisotropies}
\par
Density perturbations give rise to anisotropies in the temperature of the
CMB. There are three main contributions (see Fig. 19):
\item
{i)} Density fluctuations at the time of last scattering $t_{rec}$
imply that the surface of constant temperature $T_{rec}$ was not flat.
Hence, photons arriving at the observer $O$ from different directions
(separated by angle $\theta$) have travelled a different amount of time.
Since they were emitted with the same frequency, they arrive with differing
frequencies, thus leading to temperature anisotropies. This is the
Sachs-Wolfe effect $^{109)}$.
\item
{ii)} A localized clump of energy between last scattering and the present time
will distort the geodesics passing through it, and hence lead to
temperature differences (the Rees-Sciama effect $^{110)}$).
\item
{iii)} If either the last scattering surface or the observer are moving with
respect to the frame given by the background cosmology, Doppler
anisotropies will result.

\goodbreak \midinsert \vskip 4.5cm
\hsize=6in \raggedright
\noindent {\bf Figure 19.}
Space-time plot sketching the origin of CMB temperature anisotropies.
The surface labelled $T_{rec}$ is the last scattering surface.
$O$ is the observer at the present time measuring photons $\gamma $ impinging
from directions in the sky separated by angle $\vartheta $.
The shaded area labelled $C$ stands for a local overdensity, leading to
distortions of geodesics.
Possible velocities of observer and emitter are indicated as $\vec{v_o} $ and
$\vec{v_e}$.
\endinsert

The observed $^{111)}$ dipole anisotropy of the CMB is used to determine
the Earth's velocity relative to the rest frame of the CMB. After
subtracting the dipole, the remaining anisotropy on large angular scales
is dominated by the Sachs-Wolfe effect (see e.g. Ref. 30). The specific
relationship between $\delta T/T$ and density perturbations in the case of
adiabatic fluctuations is given by $^{30,104)}$
$${\delta T\over T}(t_o,\vartheta )={1\over 3}\Phi \left( t_{rec},\lambda
(\vartheta )\right) . \eqno\eq $$
According to relativistic perturbation theory, the relativistic potential
$\Phi $ of (5.9) is independent of time in the absence of changes in
the equation of state of the background $^{104)}$. Hence

$${\delta T\over T} (t_o , \vartheta )=  {1\over 3} \Phi \left( t_H  ( \lambda
),\lambda (\vartheta
)\right) \simeq {1\over 3}{\delta\rho\over\rho }\left( t_H (\lambda ),\lambda
\right) \eqno\eq $$
where $\lambda (\vartheta)$ is the comoving length scale corresponding to
angular separation $\theta$ at $t_{rec}$, and $t_{H}(\lambda)$ is the
time when this scale equals the Hubble radius $t$. We have used the fact
that at Hubble radius crossing $\Phi$ and $\delta \rho/\rho$ coincide up
to a numerical factor of the order 1.$^{104)}$

Recently, both the COBE satellite $^{11)}$ and MIT balloon$^{112)}$ experiment
have
detected CMB anisotropies on angular scales larger than $7^{\circ}$. The
quadrupole anisotropy reported is
$${\delta T\over T}(90^o)\simeq 5\cdot10^{-6} , \eqno\eq $$
and the low $l$ harmonies of $\delta T/T$ are consistent with a power law
spectrum of density perturbations with
$$n=1.1\pm 0.5\; . \eqno\eq $$
It is reassuring that the observed value of $n$ is so close to the value
$n=1$ favored in both inflationary and topological defect models of
structure formation. The amplitude of $\delta T/T$ is also of the right
order of magnitude to agree with theoretical predictions based on
structure formation arguments. Since by (5.8) density perturbations grow
as $a(t)$ and since today $\delta \rho/\rho$ on scales of clusters is
about $1$, we predict that on these scales

$${\delta T\over T}(\vartheta )\sim {1\over 3} z(t_{eq})^{-1}\sim 3\cdot
10^{-5}\; . \eqno\eq $$
This result agrees with the extrapolation of the COBE results to cluster
scales $^{11)}$ to better than a factor of $2$.

The results (5.24) and (5.25) cannot be used to differentiate between
inflationary Universe and topological defect theories, since the models all
predict a similar slope of $P(k)$.  However, the amplitude of $\delta
T/T$ can be used to normalize the power spectrum in any given model.
Large scale structure observations provide an independent
normalization.  These two normalizations must be consistent in order
for a theoretical model to work.  For cosmic strings, the
normalizations of $P(k)$ agree$^{106,108)}$ well, for an unbiased
texture model, the normalization factors differ by about 3$^{107)}$.

In the near future, maps of CMB anisotropies will be obtained$^{113)}$
which are signal dominated in every pixel (the COBE maps are dominated
by noise).  At that point, statistical$^{114,115)}$ measures of CMB
maps can be evaluated which can pick out the non-Gaussian
signatures of topological defect models.  Indeed, non-random phases of
the Fourier modes of $\delta \rho$ are the key feature of such models
(see Fig. 20).
\midinsert \vskip 4.8cm
\hsize=6in \raggedright
\noindent {\bf Figure 20.} A comparison of density perturbations along
a line in space between (a) random phase models and (b) theories based
on nonadiabatic seeds. \endinsert

Note that CMB anisotropies form a more severe testing ground for
cosmological models than large-scale structure data obtained from optical
telescopes because they are (at least an angular scales larger than the
horizon at last scattering) not contaminated by hydrodynamical and
nonlinear effects.

\section{Large-Scale Structure Data}
\par
Here, I shall briefly mention some of the data obtained from optical
telescopes
which at the present time seems
most relevant to cosmology.
\par
First, however, it is important to note a discrepancy between theory and
observations.  Observations become increasingly uncertain on increasing length
scales, whereas theoretical predictions decrease in accuracy as the length
scale decreases, the reason being that on large scales linear theory is
applicable and gravity is the only force which needs to be considered, whereas
on smaller scales nonlinear and messy nongravitational effects become
important.  However, this state of affairs also implies that progress either
on the theoretical or observational front will yield double benefit.  For
example, new data on larger scales not only increases the amount of data
available, but also leads into a regime with smaller theoretical
uncertainties.  Since we know that observers will be providing lots of new
data in the next few years, we can be sure that cosmology will remain an
extremely exciting field.
\par
There is a lot of new data on the large-scale structure of the Universe.
Redshift surveys have provided three dimensional maps of the distribution of
galaxies.  An example is the recent Center for Astrophysics (CFA)
survey$^{9)}$ of slices on the northern celestial sphere which show
overdense sheets of galaxies with dimensions $(50 \times 50 \times
5)$Mpc$^3$ (with $h =1$) separated by large voids.  There is evidence for
superclusters$^{116)}$, filaments longer than $50h^{-1}$Mpc$^{117)}$, voids of
diameter $60h^{-1}$Mpc$^{118)}$ and overdense sheets of
galaxies$^{119)}$ in
regions of the sky different from that covered in the CFA survey.
\par
A second window on the large-scale structure of the Universe comes from
measuring the peculiar velocities of galaxies averaged over large
regions$^{120)}$ from which one can infer information about the magnitude of
density perturbations on large scales$^{121)}$.  Preliminary observations
indicate surprisingly large velocities on scales of $60h^{-1}$Mpc, although
consensus on the interpretation of the data is still
lacking$^{120,122)}$.
\par
On the scale of clusters the data is still rather uncertain.  The quantities
one would like to focus on is the overall mass scale (the mean mass of an
object which satisfies a fixed operational definition of a cluster), the
multiplicity function and the two point (and in the future higher point)
correlation functions.
\par
A cluster of galaxies (Abell$^{123)}$ cluster) is defined as a region in the
sky with more than 50 bright galaxies in a sphere of radius $1.5h^{-1}$Mpc.
(To be compared to the mean separation of bright galaxies$^{124)}$ which is
about $5h^{-1}$Mpc).  The mean separation of clusters is about $50 h^{-1}$Mpc,
and their masses are of the order
$10^{14}$M$\odot$~$^{125)}$.
\par
The multiplicity function $n(M)$ of clusters gives the number density of
clusters of mass $M$ per unit physical volume.  $n(M) dM$ is the number
density of objects in the mass interval $[M, M+dM]$.
\par
The two point correlation function $\xi (r)$ of clusters measures the
nonrandomness of the distribution of clusters and is defined by
$$
\xi (r) = \, < {n (r) - n_0\over{n_0}} > \, , \eqno\eq
$$
where $n(r)$ is the number density at a fixed distance $r$ from a given object
and $n_0$ is the average density, and pointed brackets indicate
averaging over the objects.
\par
The initial observational results $^{10)}$ (Figure 24) gave
$$
\xi (r) \simeq \left({r\over{r_0}}\right)^{-1.8}\eqno\eq
$$
with correlation length $r_0 \sim 25 h^{-1}$Mpc.  Recently$^{126)}$, some
criticism of these results has been raised.  However, it is unlikely that the
entire effect is fictitious.
\midinsert \vskip 10cm
\hsize=6in \raggedright
\noindent {\bf Figure 21.}  The observed cluster correlation function plotted
against relative separation
(from Ref. 10).  The solid circles are the data points.  Shown is the fit to
a $r^{-2}$ slope.
\endinsert
\midinsert \vskip 8cm
\hsize=6in \raggedright
\noindent {\bf Figure 22.}  The mass function of galaxies (determined from the
luminosity function assuming constant mass to light ratio) from Refs.
127 ((a)
from Bahcall, (b) from Binggeli).
\endinsert
\par
For galaxies, similar quantities can be measured and compared with
observations.  Large spiral galaxies have masses $M \sim 10^{11}M \odot$.
Their mean separation is about $5 h^{-1}$Mpc.  The mass
function$^{127)}$
(Figure 22) gives the distribution of masses, and the two point correlation
function describes the nonrandomness of the distribution.  The two
point correlation function of galaxies has the same power law form as
(5.28), with correlation length $r_0 \sim 8$Mpc.
For galaxies, one
can in addition probe the interior structure by measuring the velocity
rotation curve $v(r)$ which is related to the density profile (baryonic plus
dark matter) by
$$
v (r) \sim r^2 \rho (r)\, .\eqno\eq
$$
Observations$^{128)}$ indicate that far beyond the disk radius, $v (r) \sim
{\rm const}$ (Figure 23) which indicates the existence of dark matter with a
density profile
$$
\rho (r) \sim r^{-2}\, . \eqno\eq
$$
We can also measure the angular momentum of galaxies.  Typical numbers for
large spirals are in the range$^{129)}$ $10^{73} - 10^{75} cm^2 g\, s^{-
1}$.
\midinsert \vskip 5.5cm
\hsize=6in \raggedright
\noindent {\bf Figure 23.}  A typical velocity rotation curve (for NGC488 HI
data, taken from Ref. 128).  The radius is in $kpc$, the velocity in $kms^{-
1}$.
\endinsert

\chapter{Cosmic Strings and Structure Formation}

Starting point of the structure formation scenario in the cosmic
string theory is the scaling solution for the cosmic string network,
according to which at all times $t$ (in particular at $t_{eq}$, the
time when perturbations can start to grow) there will be a few long
strings crossing each Hubble volume, plus a distribution of loops of
radius $R \ll t$ (see Fig. 24).
\midinsert \vskip 6.5cm
\hsize=6in \raggedright
\noindent {\bf Figure 24.}  Sketch of the scaling solution for the
cosmic string network.  The box corresponds to one Hubble volume at
arbitrary time $t$. \endinsert

\par
The cosmic string model admits three mechanisms for structure
formation:  loops, filaments and wakes.  Cosmic string loops have the same time
averaged
field as a point source with mass$^{130)}$
$$
M (R) = \beta R \mu \, , \eqno\eq
$$
$R$ being the loop radius and $\beta \sim 2 \pi$.  Hence, loops will be seeds
for spherical accretion of dust and radiation.$^{80)}$

For loops with $R \leq t_{eq}$, growth of perturbations in a model
dominated by cold dark matter starts at $t_{eq}$.  Hence, the mass at
the present time will be
$$
M (R, \, t_0) = z (t_{eq}) \beta \, R \mu \, . \eqno\eq
$$

In the original cosmic string model$^{80,131)}$ it was assumed
that loops dominate over wakes.  In this case, the theory could be
normalized (i.e., $\mu$ could be determined) by demanding that loops
with the mean separation of clusters $d_{cl}$ (from the discussion in
Section 4.4 it follows that the loop radius $R (d_{cl})$ is determined
by the mean separation) accrete the correct mass, i.e., that
$$
M (R (d_{cl}), t_0) = 10^{14} M_{\odot} \, . \eqno\eq
$$
This condition yields$^{131)}$
$$
\mu \simeq 10^{36} {\rm GeV}^2 \eqno\eq
$$
Thus, if cosmic strings are to be relevant for structure formation,
they must arise due to symmetry breaking at energy scale $\eta
\simeq 10^{16}$GeV.  This scale happens to be the scale of unification
of weak, strong and electromagnetic interactions.  It is tantalizing
to speculate that cosmology is telling us that there indeed was new
physics at the GUT scale.
\midinsert \vskip 3.5cm
\hsize=6in \raggedbottom
\noindent{\bf Figure 25.} Sketch of the mechanism by which a long
straight cosmic string moving with velocity $v$ in transverse
direction through a plasma induces a velocity perturbations $\Delta v$
towards the wake. Shown on the left is the deficit angle, in the
center is a sketch of the string moving in the plasma, and on the
right is the sketch of how the plasma moves in the frame in which the
string is at rest.
\endinsert
\par
The second mechanism involves long strings moving with relativistic
speed in their normal plane which give rise to
velocity perturbations in their wake.  The mechanism is illustrated in
Fig. 25:
space normal to the string is a cone with deficit angle$^{132)}$
$$
\alpha = 8 \pi G \mu \, . \eqno\eq
$$
If the string is moving with normal velocity $v$ through a bath of dark
matter, a velocity perturbation
$$
\delta v = 4 \pi G \mu v \gamma \eqno\eq
$$
[with $\gamma = (1 - v^2)^{-1/2}$] towards the plane behind the string
results$^{133)}$.  At times after $t_{eq}$, this induces planar overdensities,
the most
prominent (i.e. thickest at the present time) and numerous of which were
created at $t_{eq}$, the time of equal matter and
radiation$^{134,135)}$.  The
corresponding planar dimensions are (in comoving coordinates)
$$
t_{eq} z (t_{eq}) \times t_{eq} z (t_{eq}) v \sim (40 \times 40 v) \,
{\rm Mpc}^2
\, . \eqno\eq
$$

An intuitive understanding of the origin of the above distinguished scale
can be obtained as follows. Viewed from a distance, the density
perturbation grows as in a linear theory i.e.
$$ {\delta\rho\over\rho }(t)=\left( {t\over t_i }\right)^{2/3}
{\delta\rho\over\rho }(t_i) \eqno\eq $$
for a perturbation set up at a time $t_{i} > t_{eq}$. Since the initial
$\delta \rho/\rho (t_i)$ is independent of $t_{i}$, the largest density
contrast comes from the earliest $t_{i}$, namely $t_{i} = t_{eq}$.
Fluctuations created at $t_{i} < t_{eq}$ are erased by the large thermal
velocities. Thus, strings at $t_{eq}$ create the most prominent wakes.
They are also the must numerous, since the comoving separation of strings
decreases as $t_{i}$ decreases.

A more rigorous way to obtain the above result is to consider the
evolution of the velocity perturbations induced by a wake in the
Zel'drovich approximation $^{30)}$. The height of a dark matter particle
above the wake can be written as
$$h(q,t)=a(t)[q-\psi (q,t)]\; ,\eqno\eq $$
where $q$ is the initial comoving distance, and $\psi (q,t)$ is the
comoving displacement caused by the presence of the wake. The thickness of
the wake at time $t$ is determined by the value of $q$ for which
$$\dot{h} (q,t)=0\; .\eqno\eq $$
Obviously, the value of $q$ for which $\dot{h}=0$ increases as the time when
the
perturbation $\psi$ begins to grow is moved back in time. Hence,
the earliest wakes will be the thickest.

The details of the calculation depend on whether the dark matter is hot or
cold (see Chapter 8). For hot dark matter, the large thermal velocities delay
the beginning of the growth of $\psi$ on small scales. A detailed
analysis $^{136,137)}$ shows that for hot dark matter no perturbations become
nonlinear unless
$$G \mu >5\cdot10^{-7}\eqno\eq $$
In this case, the value of the redshift $z(q)$ at which $\dot{h}=0$ is
maximal for the value of $q$ (the thickness) given by
$$d\sim G \mu v\;\gamma (v)\;z(t_{eq})^2t_{eq}\sim 4v\;{\rm Mpc} \eqno\eq $$
for wakes created at $t_{i}=t_{eq}$. Note that the scales of cosmic string
wakes (see (6.7) and (6.12)) compare favorably with the measures of the
observed sheets of galaxies $^{9)}$.

Wakes arise if there is little small scale structure on the string.
In this case, the string tension equals the mass density, the string
moves at relativistic speeds, and there is no local gravitational
attraction towards the string.

In contrast, if there is small scale structure on strings,
then$^{95)}$ the string tension $T$ is smaller than the mass per unit
length $\mu$ and the metric of a string in $z$ direction becomes
$$
ds^2 = (1 + h_{00}) (dt^2 - dz^2 - dr^2 - (1 - 8G \mu) r^2 d\theta^2 )
\eqno\eq
$$
with
$$
h_{00} = 4G (\mu - T) \ln \, {r\over r_0} \, , \eqno\eq
$$
$r_0$ being the string width.  Since $h_{00}$ does not vanish, there
is a gravitational force towards the string which gives rise to
cylindrical accretion, thus producing filaments.

As is evident from the last term in the metric (6.13), space
perpendicular to the string remains conical, with deficit angle given
by (6.5).  However, since the string is no longer relativistic, the
transverse velocities $v$ of the string network are expected to be
smaller, and hence the induced wakes will be shorter and thinner.

Which of the mechanisms -- filaments or wakes -- dominates is
determined by the competition between the velocity induced by $h_{00}$
and the velocity perturbation of the wake.  The total velocity
is$^{138, 139)}$
$$
u = - {2 \pi G (\mu - T)\over{v \gamma (v)}} - 4 \pi G \mu v \gamma
(v) \, , \eqno\eq
$$
the first term giving filaments, the second producing wakes.  Hence,
for small $v$ the former will dominate, for large $v$ the latter.

By the same argument as for wakes, the most numerous and prominent
filaments will have the distinguished scale
$$
t_{eq} z (t_{eq}) \times d_f \times d_f \eqno\eq
$$
where $d_f$ can be calculated using the Zel'dovich approximation.

\chapter{Textures and Structure Formation}

Starting point of the texture scenario of structure formation is the
scaling solution for textures: at any time $t$, there is a fixed
probability $p$ that a texture configuration is entering the Hubble
radius and starting to collapse.

In the texture model it is the contraction of the field configuration which
leads to density perturbations.  At the time when the texture enters the
horizon, an isocurvature perturbation is established:  the energy density in
the scalar field is compensated by a deficit in radiation.  However, the
contraction of the scalar field configuration leads to a clumping of gradient
and kinetic energy at the center of the texture$^{140)}$ (Fig. 26).  This, in
turn, provides the
seed perturbations which cause dark matter and radiation to collapse in a
spherical manner$^{140-143)}$.

\midinsert \vskip 5.5cm
\hsize=6in \raggedbottom
\noindent {\bf Figure 26}: A sketch of the density perturbation produced
by a collapsing texture.  The left graph shows the time evolution of
the field $\chi (r)$ as a function of radius $r$ and time (see
(3.55)).  The contraction of $\chi (r)$ leads to a spatial gradient
energy perturbation at the center of the texture, as illustrated on
the right.  The energy is denoted by $\rho$.  Solid lines denote the
initial time, dashed lines are at time $t + \Delta t$, and dotted
lines correspond to time $t + 2 \Delta t$, where $\Delta t$ is a
fraction of the Hubble expansion time (the typical time scale for the
dynamics).
\endinsert

The texture model has a significant advantage over the cosmic string
theory: it is much more amenable to analytical and numerical calculations.
The reason is simple: whereas the evolution of the network of cosmic
stings is very complicated, and the detailed statistics of the scaling
solution are not yet known with good accuracy, the evolution of textures is
essentially trivial: they collapse in less than an Hubble expansion time
and set up well defined cosmological perturbations whose subsequent
evolution can be analyzed without great problems.

In the scaling solution, the comoving number density of textures $n$
obeys$^{8,69)}$.
$${dn\over d\tau }={c\over \tau^4}\eqno\eq $$
where $c \simeq 0.04$ (see also Refs. 97 and 100), and $\tau$ is
conformal time determined by
$$dt=a(t)d\tau\; .\eqno\eq $$
A collapsing texture with center at $\underline{x} = 0$ induces a velocity
perturbation$^{8,140)}$.
$$\Delta\underline{v} =-F(\underline{x},\tau
)\varepsilon\underline{\hat{x}}\eqno\eq $$
where $\underline{\hat{x}}$ is the unit vector in direction
$\underline{x}$,
$$\varepsilon = 8\pi^2G\eta^2 \eqno\eq $$
is the measure of the strength of the velocity kick (recall that $\eta$ is
the symmetry breaking scale, the single free parameter in the texture
model) and,
$$F(\underline{x},\tau )\simeq\cases{(1-{x\over\tau }) & $x<\tau $ \cr 0&
$x>\tau $} \eqno\eq $$
gives the profile of the velocity field. In the above $\underline{x}$ is a
comoving coordinate. In agreement with the causality conditions, $F$
vanishes on scales larger than the horizon $(|\underline{x}|\equiv x >
\tau)$.

The density fluctuation $\delta (\underline{x}, \tau, \tau_{i})$ at time
$\tau$ induced by a texture collapsing at $x=0$ at time $\tau_{i}$ is
determined by the initial velocity perturbation. The result is $^{8,141)}$
$$\delta (x,\tau ,\tau_i)={2F(\underline{x},\tau_i )\varepsilon\tau_*
\over x}a(\eta_i)[\delta_2(\eta_i)\delta_1(\eta )-\delta_1(\eta_i)\delta_2(\eta
)]
\eqno\eq $$
where  $\delta_{1} (\eta)$ and $\delta_{2}(\eta)$ are the growing and
decaying mode solutions of the linear perturbation equations, and
$${1\over\tau_* }=\left( {8\pi G\rho (t_{eq} )\over 3}\right)^{1/2}\;.\eqno\eq
$$
In the above, the scale factor has been normalized such that
$$a(t_{eq} )=1\; .\eqno\eq $$
In this case the growing mode is
$$\delta_1(a(\eta ))=1+{3a\over 2}\; .\eqno\eq $$

Based on (7.1) and (7.6), many interesting observables can be computed
rather easily. As an example, in Fig. 27 the resulting power spectrum of
density perturbations is shown and compared to what is obtained in the
"standard CDM model", a theory based on quantum fluctuations from inflation
in a Universe dominated by cold dark matter.
\goodbreak \midinsert \vskip 4.5cm
\hsize=6in \raggedright
\noindent {\bf Figure 27.}
A comparison of the power spectrum of density perturbations in a global texture
model (solid line) and in the standard CDM model (dashed curve). Note the
larger
power in the texture model on large scales (from Ref. 143). The units of $P(k)$
are arbitrary.
\endinsert

As in the cosmic string model, also in the global texture scenario the
length scale of the dominant structures is the comoving Hubble radius
at $t_{eq}$.  Textures generated at $t_{eq}$ are the most numerous,
and the perturbations induced by them have the most time to grow.

In both cosmic string and texture models, the fluctuations are non-Gaussian,
which means that the Fourier modes of the density perturbation $\delta \rho$
have nonrandom phases.  Most inflationary Universe models, in contrast,
predict (in linear theory) random phase fluctuations which can be
viewed as a superposition of small amplitude plane wave perturbations with
uncorrelated phases (for some subtle issues see Refs. 144 and 145).

\chapter{Comparison}
\section{Role of Hot and Cold Dark Matter}
\par
Before discussing some key observations which will allow us to distinguish
between the different models, I will discuss the role of dark matter.  The key
issue is free streaming.  Recall that cold dark matter consists of particles
which have negligible velocity $v$ at $t_{eq}$, the time when sub-horizon
scale perturbations can start growing:
$$
v (t_{eq}) \ll 1 \qquad {\rm (CDM)} \, . \eqno\eq
$$
For hot dark matter, on the other hand:
$$
v (t_{eq}) \sim 1 \qquad {\rm (HDM)} \, . \eqno\eq
$$
Due to their large thermal velocities, it is not possible to establish
HDM perturbations at early times on small scales.  Fluctuations are
erased by free streaming on all scales smaller than the free streaming
length
$$
\lambda_J^c (t) = v (t) z (t) t \eqno\eq
$$
(in comoving coordinates).  For $t > t_{eq}$, the free streaming
length decreases as $t^{-1/3}$.  The maximal streaming length is
$$
\lambda_J^{\rm max} = \lambda^c_J (t_{eq}) \eqno\eq
$$
which for $v (t_{eq}) \sim 0.1$ (appropriate for 25 eV neutrinos)
exceeds the scale of galaxies.
\par
In inflationary Universe models and in the texture theory, the density
perturbations are essentially dark matter fluctuations.  The above free
streaming analysis then shows that, if the dark matter is hot, then
no perturbations on the scale of galaxies
will survive independent of larger-scale structures.  Hence, these
theories are acceptable only if most of the dark matter is cold.
\par
Cosmic string theories, in contrast, work well - if not even better -
with hot dark matter$^{146,147,7,136,137)}$.  The cosmic string seeds survive
free
streaming.  The growth of perturbations on small scales $\lambda$ is
delayed (it starts once $\lambda = \lambda_J (t)$) but not
prevented.
\par
Let us summarize the main characteristics of the cosmic string, global
texture, and inflationary Universe theories of structure formation.
Inflation predicts random phase perturbations.  The density peaks will
typically be spherical, and the model is consistent with basic
observations only for CDM.  The global texture and cosmic string
models both give non-random phase perturbations.  The topology is
dominated by spherical peaks for textures, whereas it is planar or
filamentary for
cosmic strings (depending on the small scale structure on the
strings).  Textures require CDM, whereas cosmic strings work
better with HDM.

\section{Large-Scale Structure Signatures for Cosmic Strings and
Textures}
\par
The \underbar{genus curve}$^{148)}$ is a statistical measure for the topology
of
large-scale structure.  Given a smooth density field $\rho
(\undertext{x})$, we pick a density $\rho_0$ and consider the surface
$S_{\rho_0}$ where $\rho (\undertext{x}) = \rho_0$ and calculate the
genus $\nu (S)$ of this surface
$$
\nu =  {\rm \# \> of \> holes \> of \>} S - {\rm \# of \> disconnected
\> components \> of \>} S \, . \eqno\eq
$$
The genus curve is the graph of $\nu$ as a function of $\rho_0$.
\par
For perturbations with Poisson statistics, the genus curve can be
calculated analytically (Fig. 28).  The inflationary CDM model in the
linear regime falls in this category.  The genus curve is symmetric
about the mean density $\bar \rho$.  In the texture model, the
symmetry about $\bar \rho$ is broken and the genus curve is shifted to
the left$^{143)}$.  In the cosmic string model, there is a pronounced
asymmetry between $\rho > \bar \rho$ and $\rho < \bar \rho$.  At small
densities, the genus curve measures the (small number) of large voids,
whereas for $\rho > \bar \rho$ the curve picks$^{149)}$ out the large
number of high density peaks which result as a consequence of the
fragmentation of the wakes (Fig. 28).

\midinsert \vskip 10cm
\hsize=6in \raggedbottom
\noindent{\bf Figure 28.} The genus curve of the smoothed mass density
field in a cosmic string wake toy model compared to the symmetric
curve which results in the case of a model with a random distribution
of mass points. The vertical axis is the genus (with genus zero at the
height of the ``x"), the horizontal axis is a measure of density (``0"
denotes average density).
\endinsert

The \undertext{counts in cell statistic}$^{150)}$ can be successfully
applied to distinguish between distributions of galaxies with the same
power spectrum but with different phases.  The statistic is obtained
by dividing the sample volume into equal size cells, counting the
number $f(n)$ of cells containing $n$ galaxies, and plotting $f(n)$ as
a function of $n$.

We$^{151)}$ have applied this statistic to a set of toy models of
large-scale structure.  In each case, the sample volume was (150
Mpc)$^3$, the cell size (3.75 Mpc)$^3$, and the samples contained
90,000 galaxies.  We compared a texture model (galaxies distributed in
spherical clumps separated by 30 Mpc with a Gaussian radial density
field of width 9 Mpc), a cosmic string model dominated by filaments
(all galaxies randomly distributed in filaments of dimensions (60
$\times$ 4 $\times$ 4) Mpc$^3$ with mean separation 30 Mpc, a cosmic
string wake model (same separation and wake dimensions $(40 \times 40
\times 2)$ Mpc$^3$), a cold dark matter model (obtained by Fourier
transforming the CDM power spectrum and assigning random phases), and
a Poisson distribution of galaxies.

\midinsert \vskip 17cm
\hsize=6in \raggedbottom
\noindent{\bf Figure 29.}
The three dimensional counts in cell statistics for a Poisson model
(G), a cold dark matter model (CDM), cosmic string wakes (SW), string
filaments (SF) and textures (T).
\endinsert

As shown in Fig. 29, the predicted curves differ significantly,
demonstrating that this statistic is an excellent one at
distinguishing different theories with the same power spectrum.  The
counts in cell statistic can also be applied to effectively two
dimensional surveys such as single slices of the CFA redshift
survey$^{9)}$.  The predictions of our theoretical toy models are
shown in Fig. 30.

\midinsert \vskip 17cm
\hsize=6in \raggedbottom
\noindent{\bf Figure 30.}
The two dimensional counts in cell statistic for a slice of the
Universe of the dimensions of a CFA slice, evaluated for the same
models as in Fig. 32.
\endinsert

A third statistic which has proved useful$^{152)}$ is distinguishing
models with Gaussian and non-Gaussian phases is the void probability
function $p(R)$, the probability that a sphere of radius $R$ contains
no galaxies.

\section{Signatures in the Microwave Background}

Inflationary Universe models predict essentially random phase
fluctuations in the microwave background with a scale invariant
spectrum $(n = 1)$.  Small deviations from scale invariance are model
dependent and have recently been discussed in detail in Refs. 105.  In all
models, the amplitude must be consistent with
structure formation. COBE
discovery$^{11)}$ of anisotropies in the CMB has provided severe
constraints on inflationary models.  They are only consistent with the
present data if the bias parameter $b$ is about 1, which must be
compared to the value $b \simeq 2.5$ which is the best value for
galaxy formation in this model$^{153)}$.  Note that full sky coverage
is not essential for testing inflationary models since in any set of
local observations of $\delta T/T$, the results will form a Gaussian
distribution about the r.m.s. value.  Mixed dark matter models do
slightly better and have recently been studied vigorously.

\midinsert \vskip 7cm
\hsize=6in \raggedbottom
\noindent{\bf Figure 31.}
Sketch of the mechanism producing linear discontinuities in the
microwave temperature for photons $\gamma$ passing on different sides
of a moving string $S$ (velocity $v$).  $O$ is the observer.
Space perpendicular to the string is conical (deficit angle $\alpha$).
\endinsert

Cosmic string models predict non-Gaussian temperature anisotropies.
One mechanism gives rise to localized linear temperature
discontinuities$^{154)}$; its origin is illustrated in Fig. 31.  Photons
passing on different sides of a long straight string moving with
velocity $v$ reach the observer with a Doppler shift
$$
{\delta T\over T} \sim 8 \pi G \mu v \gamma (v) \, . \eqno\eq
$$
To detect such discontinuities, an appropriate survey strategy ({
e.g.}, full sky survey) with small angular resolution is crucial.  The
distribution of strings also gives rise to Sachs-Wolfe type
anisotropies$^{155)}$.

The theoretical error bars in the normalization of CMB anisotropies
from strings are rather large -- a direct consequence of the fact that
the precise form of the scaling solution for the string network is not
well determined.  Nevertheless, we can consider a fixed set of cosmic
string parameters and ask whether the normalizations of $G\mu$ from
large-scale structure data and from COBE are consistent.  This has
been done numerically in Ref. 106, and using an analytical toy model in
Ref. 108.

The analytical model$^{108)}$ is based on adding up as a random walk the
individual Doppler shifts from strings which the microwave photons
separated by angular scale $v$ pass on different sides, and
using this method to compute $\Delta T/T (\theta)$.  Using the
Bennett-Bouchet$^{156)}$ string parameters, the result for $G \mu$
becomes
$$
G\mu = (1.3 \pm 0.5) 10^{-6} \, ,  \eqno\eq
$$
in good agreement with the requirements from large-scale structure
formation$^{7)}$.

To detect the predicted anisotropies from textures, it is again
essential to have a full sky survey.  However, angular
resolution is adequate this time, since the specific signature for
textures is a small number $( \sim 10)$ of hot and cold disks with
amplitude$^{157)}$
$$
{\delta T\over T} \sim 0.06 \times 16 \> \pi \> G \eta^2 \sim 3 \cdot 10^{-5}
\eqno\eq
$$
and angular size of about $10^\circ$.
The hot and cold spots are due to photons
falling into the expanding Goldstone boson radiation field which
results after texture collapse or due to photons climbing out of the
potential well of the collapsing texture$^{158)}$ (see Fig. 32).

\midinsert \vskip 5cm
\hsize=6in \raggedbottom
\noindent{\bf Figure 32.} Space-time diagram of a collapsing texture
(backward light cone) and the resulting expanding Goldstone boson
radiation (forward light cone). Unwinding of the texture occurs at
point ``TX". The light ray $\gamma_2$ falls into the potential well
and is blueshifted, the ray $\gamma_1$ is redshifted.
\endinsert

Note that the texture model is not ruled out by the recent COBE
results.  The amplitude (8.8) is lower than the pixel sensitivity of
the COBE maps.  However, the predicted quadrupole CMB anisotropy
(normalizing $\eta$ by the large-scale structure data) exceeds the
COBE data point by a factor of between 2.5 and 4$^{107,159)}$.  Hence,
biasing must be invoked in order to try to explain the large-scale
structure data given the reduced value of $\eta$ mandated by the
discovery of CMB anisotropies.

\section{Conclusions}

Topological defect models of structure formation generically give rise
to a scale invariant power spectrum and are hence in good agreement
with the recent COBE results on anisotropies of the CMB.  The
amplitude of the quadrupole temperature fluctuation can be used to
normalize the models.  For cosmic strings, the resulting normalization
agrees well with the normalization from large-scale structure data.
For textures, there is a mismatch which requires introducing biasing.
For textures, the situation is comparable to that in the CDM model,
where COBE demands a bias parameter $b \simeq 1$, whereas galaxy
formation is said to demand$^{153)}$ $b \simeq 2.5$.

It was emphasized that r.m.s. data intrinsically is unable to
differentiate between topological defect models (with non-random
phases) and inflationary modes (with random phases).  We need
statistics which are sensitive to nonrandom phases.

The most economical model for structure formation may be the model
based on cosmic strings and hot dark matter.  It requires no new
particles (although it does require a finite neutrino mass), it agrees
well with COBE and with the CFA redshift data, and it has clear
signatures both for large-scale structure and CMB statistics.

\chapter{Microphysics of topological defects}

In Chapters 5-8, applications of topological defects were considered for
which only the gravitational effects of such defects were important.
However, there is much more structure in defects than is visible through
gravity only. Topological defects are localized coherent excitations of
gauge and Higgs fields. Matter interactions with the defects can probe
this structure, can give rise to effects of potential importance in
cosmology, and possibly even to new direct signatures for defect models.
In this and the following chapter, I will briefly touch on two aspects
involving microphysics. The discussion is not intended to be complete, but
rather wet the curiosity of the reader. First, we consider scattering of
fermions by strings and monopoles.

\section{Callan-Rubakov Type Effects}

Consider cosmic strings arising in a GUT model. The gauge and Higgs fields
excited in the string mediate baryon and lepton number violating
interactions, e.g. (in the case of the Higgs field $\phi$) via a term
$ {\cal L}_{I }$ in the interaction Lagrangian
$$ {\cal L}_{I }=g\overline{\psi }\phi\psi\; ,\eqno\eq $$
where $g$ is the Yukawa coupling constant and $\psi$ is a $G$ multiplet of
fermion fields, $G$ being the gauge group.

Due to the term (9.1) in the interaction Lagrangian (and similar terms due
to the GUT gauge fields coupling to $\psi$) there is a nonvanishing
cross section $\sigma$ for a quark to be scattered into a lepton by the
defect. The conserved charges which are different in the initial and final
state are absorbed by the defect (which is here treated as a static
background field).

In the case of fermion-monopole scattering, we would naively expect the
cross section to be the geometrical one $\sigma_{geom}$
$$\sigma_{geom}\sim\eta^{-2}\; ,\eqno\eq$$
$\eta^{-1}$ bein the monopole radius. However, as first discovered by
Callan $^{160)}$ and Rubakov $^{161)}$, there is in fact an enhancement of the
cross section and
$$\sigma\sim m^{-2}\; ,\eqno\eq $$
$m$ being the fermion mass. Note that (9.3) represents a strong
interaction cross section. Is there a similar enhancement for cosmic
strings?

Classical physics considerations indicate that there will be no enhancement
of the baryon number violating inelastic cross section for fermion-cosmic
string scattering $^{162)}$. For monopoles, the enhancement of the cross
section can be viewed $^{162)}$ as the consequence of an attractive magnetic
moment-magnetic field force for $s$ wave scattering (see Fig. 33). The
potential energy is
$$V(r)\sim -(\underline{\mu }\cdot\underline{B} )\sim {1\over r^2 }\; ,\eqno\eq
$$
where $\underline{B}$ is the  $U(1)$ magnetic field of monopole, and
$\underline{\mu}$ is the magnetic moment of the fermion. This results in
an attractive force
$$F(r)\sim {1\over r^3}\; . \eqno \eq $$
In the case of ordinary (i.e. non-superconducting) cosmic strings, there
is no long range magnetic field and hence no enhancement of the inelastic
scattering cross section due to classical physics effects $^{162)}$. For
superconducting cosmic strings, $\underline{B}$ is circular in the plane
perpendicular to the string, and  hence again $V(r)$ vanishes for $s$
wave scattering $^{163)}$.

\midinsert \vskip 5cm
\hsize=6in \raggedbottom
\noindent{\bf Figure 33.} The magnetic field ${\bf B}$ of a magnetic monopole
$M$ is parallel to the magnetic moment ${\bf \mu}$ of an incident s-wave
fermion $f$.
\endinsert

However, this is not the full story. Quantum mechanical effects due to the
nonvanishing gauge field at long distances from the string may induce an
enhancement of the cross section, an effect reminiscent but not identical
to the Aharonov-Bohm $^{164)}$ effect. (The AB effect appears in elastic
scattering whereas our effect is due to inelastic scattering). Let us
therefore consider the quantum mechanical scattering of fermions by cosmic
strings.

Fermion-cosmic string scattering was first considered using a first
quantized approach in Refs. 165 and 166; and extended to inelastic
scattering in Ref. 167. The same results can be derived $^{168)}$ using a
perturbative second quantized approach discussed in Ref. 162.

We consider fermion-defect scattering to first order in perturbation
theory, treating the defect fields as a fixed background. The transition
matrix element is
$${\cal A} ={}_{\infty }\! < \psi 'D|\psi D>_{-\infty }=<\psi 'D|S|\psi D>
\eqno\eq$$
where $D$ stands for the defect and $S$ is the scattering matrix. To first
order in $g$
$${\cal A}=<\psi 'D|\int d^4x{\cal L}_{I}|\psi D> $$
$$\;\;\; = g\int dtd^3\underline{x} <\psi '|\overline{\psi }\psi |\psi
><D|\phi|D>\; ,\eqno\eq $$
where we have split the Hilbert space expectation value into its two
tensor components.

{}From $\cal A$ we obtain the differential cross section by the usual
relation
$${d\sigma\over d\Omega }\sim {1\over T}V\!\int d^3k'|{\cal A}|^2\; ,\eqno\eq
$$
where $k'$ is the momentum of the final particle and $V,T$ are cutoff
volume and time respectively.

The procedure $^{162)}$ now is to first evaluate $\cal A$ using free fermion
wave functions, obtaining the unamplified cross section
$(d\sigma/d\Omega)_{\circ}$. In a second step, we solve the Dirac equation
in the presence of the defect and determine the amplification factor
$$A\equiv {\psi (\eta^{-1} )\over\psi_{free}(\eta^{-1} )}\eqno\eq$$
as the ratio of the actual and free wave functions $\psi$ and
$\psi_{free}$, respectively, evaluated at the core radius. Care must be
taken to normalize both wave functions in the same way at infinity. Note
that $r=0$ (or $\rho = 0)$ is taken to be the defect center for monopoles
(or strings). The actual cross section becomes
$${d\sigma\over d\Omega }=A^4\left( {d\sigma\over d\Omega }\right) _o\eqno\eq
$$
(four powers of A because in $|{\cal A}|^{2}$ two fermion wave functions
appear and both are squared).

As an example $^{168)}$, consider the defect to be a cosmic string. The
``geometric" amplitude ${\cal A}_{0}$ for a string along the $z$ axis is
$${\cal A}_o\sim g\eta\eta^{-2}\int dtdz{m\over
(E_kE_{k'})^{1/2}V^{1/2}}e^{i(E_k-E_{k'})t}e^{i(k_z'-k_z)z}\eqno\eq $$
The first factor of $\eta$ comes from the amplitude of $\phi$ inside the
core, integration over the core gives $\eta^{-2}$, the factor $m$ comes
from fermion spin sums (strictly speaking the summation only occurs when
computing the cross section), and the other factor $m/\eta$ is due to the
suppression of the free s-wave amplitude at the core radius. Taking the
square and performing one $z$ and $t$ integral gives
$$({\cal A}_o)^2\sim g^2\left( {m\over\eta }\right) ^2{1\over E_k^2V}{\rm
LT}\delta
(E_k-E_{k'})\delta (k_z-k_{z'})\left( {m\over\eta }\right) ^2\eqno\eq $$
The two delta functions express conservation of fermion energy and
momentum along the string. $L$ is a cutoff length. Integrating over
$\underline{k'}$ and taking $E_{k'} \sim m$ gives the following
differential cross section per unit length
$${d\sigma\over d\Omega d{\rm L}}\sim g^2\left( {m\over\eta }\right) ^4
{1\over m}\; .\eqno\eq $$
The second step of the analysis is to calculate the amplification of the
fermion wave function at the core radius by solving the Dirac equation in
the background field of the string:
$$i\not\!\! D\psi -m\psi =0\eqno\eq $$

The covariant derivative $D_\mu$ is
$$D_{\mu } =\partial_{\mu }+ieA_{\mu }\; .\eqno\eq $$
Here, $e$ is the charge of the fermions with respect to the subgroup of
$G$ for which $A_{\mu}$ is the gauge field. We assume that in units of
$2\pi /e$ the flux of the string is $\alpha$.

The solution of the Dirac equation makes use of the techniques pioneered by
de Vega $^{169)}$. Since we are interested in field configurations independent
of $z$, we choose a basis of $\gamma$ matrices in which only $\gamma_{z}$
mixes the upper and lower 2-spinors of the 4-spinor $\psi$.
$$\psi =\left (\matrix{\psi_u\cr\psi_d\cr }\right ) \eqno\eq $$
(see Refs. 162 and 168 for details). The Dirac equation (9.14) then
decouples into two separate equations for $\psi_{u}$ and $\psi_{d}$
respectively.
Each is a system of two coupled first order differential equations.

Next, we expand $\psi_{u}$ (and similarly $\psi_{d}$) in partial waves
$$\psi_{\mu } =\sum_{n=-\infty }^{\infty }{}\left (\matrix{F_n(\rho )\cr
G_n(\rho )e^{i\vartheta }\cr }\right ) e^{i n \vartheta -i \omega t}\; ,
\eqno\eq $$
where $\rho$ and $\vartheta$ are the polar coordinates in the plane
perpendicular to the string. The radial function $F_{n}(\rho)$ satisfies a
Bessel equation with index depending on $\alpha$. It is necessary to solve
the equation inside and outside the core, and to match the solutions at
the core radius. For simplicity, we take the flux of the string to be
concentrated on the cylinder $\rho = \eta^{-1}$.

Finally, for given $\alpha$ we search for the mode $n(\alpha)$ for which
$F_{n}(\eta^{-1})$ is largest and use this coefficient to determine the
amplification factor A. We find nontrivial amplification:
$$A=\left( {\eta\over m}\right) ^{p(\alpha )}\; ,\eqno\eq $$
where $p(\alpha)$ is plotted in Fig. 34.

\midinsert \vskip 6.5cm
\hsize=6in \raggedbottom
\noindent{\bf Figure 34.} The amplification exponent $p$ plotted as a function
of the fractional flux $\alpha$ on the string, in a model in which the fermions
do not couple to the Higgs field.
\endinsert

We conclude that in general, there is enhancement of the inelastic baryon
number violating cross section due to intrinsically quantum mechanical
effects, namely the presence of a nonvanishing gauge potential in the far
field limit. The maximal cross section is
$$\left. {d\sigma\over d\Omega d{\rm L}}\right| _{max}\sim {1\over m}\;
,\eqno\eq $$
i.e. a strong interaction cross section. However in general $d \sigma / d
\Omega d {\rm L}$ is suppressed, the suppression rate depending on the
flux/charge ratio $\alpha$.

\section{Nonabelian Aharonov-Bohm Effect}

When fermions are scattered by a flux tube, the elastic cross section is
also much larger than the  geometric cross section, as was first discussed
by Aharonov and Bohm $^{164)}$. The scattered wave has amplitude
$$|f(\vartheta )|={1\over (2\pi k)^{1/2}}{\sin (\pi\alpha )\over\cos (\vartheta
/2)}\; ,\eqno\eq $$
$\vartheta$ being the scattering angle, and $k$ the wave number of the
fermion.

The above corresponds to scattering of a fermion by a $U(1)$ cosmic string.
It is not hard to see what happens when extending this analysis to
fermions scattering by a GUT string$^{170)}$.

Consider a GUT theory with a nonabelian string, i.e. a string in which
the generator $M$ of the gauge field $A_{\mu}$ excited in the string is
nondiagonal in the basis in which quarks and leptons are separate entries
in the fermion multiplet. As an example $^{171-174)}$, consider $G=SO(10)$ with
the
first stage of symmetry breaking
$$SO(10)\rightarrow SU(5)\times Z_2 \eqno\eq $$
giving rise to strings. The fermions are in the 16 representation of $SO
(10)$. We choose a basis in which
$$\psi =(u_1,u_2,u_3,\nu_e,d_1,d_2,d_3,e^-,d_1^c,d_2^c,
d_3^c,e^+,-u_1^c,-u_2^c,-u_3^c,-\nu_e^c )\eqno\eq $$
where $U$ and $d$ denote the usual quarks and the subscripts $1-3$ stand
for the three colors.

A string solution exists $^{174)}$ for which
$$M={1\over 2}\left (\matrix{
B & 0 & 0 & 0 \cr
0 & B & 0 & 0 \cr
0 & 0 & -B & 0 \cr
0 & 0 & 0 & -B \cr
}\right )\; ,\; B=\left (\matrix{
0 & 0 & 0 & 0 \cr
0 & 0 & 0 & 0 \cr
0 & 0 & 0 & 1 \cr
0 & 0 & 1 & 0 \cr
}\right )\; .\eqno\eq $$
This is a nonabelian string for which $A_{\mu}$ mixes quarks and leptons.
We wish to consider the elastic scattering of fermions for this string
$^{170)}$.

We may focus on a two-dimensional subspace, e.g. the $u_{3}, \nu_{e}$
subspace for which
$$ M =\pmatrix{
0 & 1 \cr
1 & 0 \cr
}\; .\eqno\eq $$
To reduce the scattering problem to that for nonabelian string we
diagonalize $M$. In the diagonal basis, the fermion eigenstates are
$$\psi_1={1\over\sqrt{2}}(u_3+\nu_e)$$
$$\psi_2 = {1\over\sqrt{2}}(u_3-\nu_e )\; .\eqno\eq $$
The charges are opposite. Hence, the scattering amplitudes differ. If
the incident beam of fermions is pure $u_{3}$, the scattered beam will be a
mixture of $u_{3}$ and $\nu_{e}$. Hence, we have a nonabelian Aharonov-Bohm
(AB) effect. The scattering violates baryon number!

The above discussion (see also Refs 175 and 176 for an $SU(2)$ example) has
been sloppy. Account must be taken of the fact that the physical basis
(i.e. the basis in which a given entry in $\psi$ has fixed physical
charges) rotates  as we go around the string. A recent careful analysis by
Ma $^{177)}$ shows that baryon number violation in the nonabelian AB effect
persists.

\chapter{Topological Defects and Baryogenesis}

Even through baryon number violation processes can be mediated by
scattering of fermions from topological defects, no net baryon asymmetry
can be created because both initial and final particles are in thermal
equilibrium. As realized already by Sakharov $^{178)}$, in order to generate a
nonvanishing baryon to entropy ratio  $n_{B}/s$, it is necessary to have
- in addition to the presence of baryon number violation processes - CP
violation, and the processes which violate baryon number must occur out of
thermal equilibrium. Two ways in which topological defects can contribute
to baryogenesis are sketched in this chapter. The first recurs at the
unification scale, the second at the electroweak scale.

\section{Collapsing Cosmic Strings and Baryogenesis}

The standard mechanism $^{179)}$ of GUT scale baryogenesis is based on the
out-of-equilibrium decay of superheavy  Higgs and gauge particles which
freeze out of the initial thermal bath of particles at a temperature
$T_{f}$ of the order of the critical temperature $T_{c}$ of the symmetry
breaking phase transition.

The above is a viable mechanism. Baryon number violating interactions
arise in the Lagrangian (see (9.1)), CP violation already exists in the
standard model, and the out of thermal equilibrium condition is achieved
as described in the previous paragraph.

However, if $T_{f}$ is significantly lower than the mass $m_{X}$ of the
superheavy fields, then the net baryon asymmetry from the above process
may be exponentially suppressed since $^{180)}$ for $T_{f} < T < m_{X}$ the
number density $n_{X}$ of $X$ particles decays exponentially
$$n_X\sim e^{-{m_X\over {\rm T}}}\; .\eqno\eq $$
As first pointed out in Refs. 181 and 182, cosmic strings can contribute to
baryogenesis in the subset of GUT models which admit cosmic string solutions.
The
topological defects form out of equilibrium concentrations of GUT Higgs
and gauge fields. Any emission of quanta form the strings proceeds through
$n_{B}$ violating interactions, and hence a net baryon number may be
generated. In Ref. 181, baryogenesis from cosmic strings collapsing onto a
line  was considered, and in Ref. 182 the contribution of cusp
annihilation $^{183)}$ to $n_{B}/s$ was calculated. A third source $^{184)}$ is
the
final collapse of a cosmic string loop.

As discussed in Chapter 4, cosmic string loops slowly shrink by emitting
gravitational radiation. Once the radius $R$ becomes comparable to the
thickness $w$ there is no topological barrier for the remnant ball of
energy
$$E\sim 2\pi\mu w\sim 2\pi\lambda ^{-1/2}\eta\eqno\eq $$
to decay into $N_{Q}$ quanta of superheavy particles which in turn decay
into low mass particles producing a net $n_{B}$. From (10.2) and $m_{X}
\sim \eta$ it follows that
$$N_Q\sim 2\pi\lambda ^{-1}\; .\eqno\eq $$
If $\Delta B$ is the average net baryon number per decay, then the net
baryon asymmetry produced per loop is
$$\Delta N_B\sim 2\pi\lambda ^{-1}\Delta B\; .\eqno\eq $$
Integrating over the distribution of loops after the time of the phase
transition and dividing by the entropy density we obtain $^{184)}$
$${n_B\over s}\sim\lambda ^3N^{-1}N_Q\Delta B\eqno\eq $$
where $\lambda$ is the self coupling constant of the Higgs field potential
and $ N$ is the number of spin degrees of freedom in thermal equilibrium.
In order to obtain this result we took $ T_{f} = T_{c}$ and the mean
separation of strings to be given by (4.9).

Since the number density of string loops decays only as a power law
(redshift effect), the contribution of collapsing cosmic string loops to $
n_{B}/s$ decays only as $T_{F}^{+3}$ and hence may dominate if $T_{F} \ll
m_{X}$.

The result (10.5) must be compared to the maximal $ n_{B}/s$ which can be
produced by the usual GUT mechanism $^{185)}$:
$${n_B\over s}\sim N^{-1}N_X\Delta B\; .\eqno\eq $$
where $N_{X}$ is the number of helicity states of all superheavy particles
contributing to $\Delta B$. Note that $\Delta B$ depends on the particular
GUT model chosen. It enters both the standard and topological defect
baryogenesis mechanism in the same way. The value of $\Delta B$ has been
estimated $^{185)}$ to lie in the range  $10^{-13} < \Delta B <10^{-2}$. Thus,
it is in principle possible to generate the observed $n_{B}/s$ ratio which
lies in the range
$${n_B\over s}\sim 10^{-10}\; . \eqno\eq $$

\section{Electroweak Baryogenesis with Electroweak Strings}

Recently, it has been realized that nonperturbative processes associated
with the $SU(2)$ anomaly $^{186)}$ of the standard model can erase a primordial
$B + L$ asymmetry above the critical temperature of the electroweak
symmetry breaking transition. Thus, unless a primordial $B - L$ asymmetry
is generated at the GUT scale (or extra symmetries such as those arising
in supersymmetry are introduced $^{187)}$), the observed $n_{B}/s$ must be
produced below the electroweak scale.

Initial ideas on how to generate a nonvanishing $n_{B}/s$ were discussed
in Ref. 188, and recently several concrete mechanisms were suggested
$^{189-191)}$. All of these mechanisms only work if the electroweak phase
transition is first order and therefore proceeds by bubble
nucleation.

In electroweak baryogenesis, $n_{B}$ violation recurs nonperturbatively
through the anomaly, CP violation is already contained in the standard
model (although stronger CP violation is introduced in Refs 189-191 by
enlarging the Higgs sector of the model). The role of the first order phase
transition in Refs. 189-191 is to generate bubble walls in which the
$n_{B}$ violating processes can take place out of thermal
equilibrium.

However, the order of the electroweak phase transition is not yet known.
Therefore it is important to point out that electroweak baryogenesis
mechanisms also exist if the phase transition is second order $^{192)}$.
Topological defects may play the role of the bubble walls in a first
order transition.

However, in the electroweak model and minimal extensions thereof such as
the two Higgs doublet model there are no stable topological defects. But,
as realized recently by Vachaspati $^{193)}$ and considered much earlier by
Nambu $^{194)}$, energetically stable string-like defects can be constructed in
the standard model by embedding the Nielsen-Olesen $U(1)$ string into the
standard model in the following way:
$$\Phi = f_{{\rm NO}}(\rho )e^{i\vartheta }\left (\matrix{ 0\cr 1\cr }\right )
\eqno\eq $$
$$Z=A_{\rm NO} \eqno\eq $$
$$W'=W^2=A=0\; , \eqno\eq $$
where $f_{{\rm NO}}(\rho)e^{i \vartheta}$ and $A_{{\rm NO}}$ are the
Nielsen-Olesen
string configurations.

It can be shown $^{195)}$ that for large Weinberg angles, the above electroweak
string configuration is energetically stable. Generally, electroweak
strings have finite length $^{193)}$. Their ends can be viewed as a
monopole-antimonopole pair.

Electroweak baryogenesis works as follows $^{189,190)}$. In the outer shell of
the
bubble wall where
$$|\Phi |<g\eta\; , \eqno\eq $$
$g$ being the gauge coupling constant, sphaleron $^{190)}$ and local texture
unwinding $^{189)}$ events take place out of thermal equilibrium at a rate
$$\Gamma_B\sim\alpha_W^4T^4\; . \eqno\eq $$
Since there is CP violation in this region, a net baryon asymmetry will be
generated. It is important that the region of $n_{B}$ violation is in
motion with a preferred direction (otherwise there is no net CP
violation).

In a model with a second order phase transition and electroweak strings,
the role of the bubble wall is played by the tips of the string which are
moving towards each other due to the tension of the string $^{192)}$.

The typical length and separation of strings immediately after the phase
transition is the Ginsburg length (see (4.9))
$$\xi (t_G)\sim\lambda^{-1}\eta^{-1}\; . \eqno\eq $$
The net baryon asymmetry produced by one electroweak string is
$$N_B=\int^{t_G+\Delta t_S}_{t_G}dt{dV\over dt}\Delta t_c\varepsilon\Gamma _B\;
,
 \eqno\eq $$
where $dV/dt$ is the rate of change of the volume in which $n_{B}$
violating processes are taking place out of equilibrium, $\Delta t_c$ is
the length of time any point in space remains in the above volume,
$\varepsilon$ is the CP violation parameter defined such that $\varepsilon
\Gamma_{B}$ is the net rate of baryon number producing processes per unit
volume. $\Delta t_S$ is the contraction time of the string.

Using (10.12) and the geometrical values for $dV/dt$,$\Delta  t_c$ and
$\Delta t_S$ (see Ref. 192 for details), we obtain
$${n_B\over s}\sim\lambda g^2\varepsilon{\alpha_W}^4\; , \eqno\eq $$
which is smaller than what is obtained with a first order phase transition
only be a geometrical factor (the region in which $n_{B}$ violation takes
place out of equilibrium is much smaller).

Note, however, that the above model of electroweak baryogenesis using
topological defects arising in a second order phase transition only works
if extreme conditions on the parameters of the theory are satisfied: we
require stable electroweak strings, a second order transition,  sufficient
CP violation $(\varepsilon \sim 1)$, constraints on the coupling constants
to make sure that sphalerons or textures fit into the region $|\phi| < g
\eta$ and that $n_{B}$ violating effects in the true vacuum are suppressed
at $t_{G} + \Delta t_{s}$. Nevertheless, it is an existence proof that
electroweak baryogenesis with a second order phase transition is possible.
\bigskip
\REF\one{ T.W.B. Kibble, {\it J. Phys.} {\bf A9}, 1387 (1976).}
\REF\thirteen{A. Guth, {\it Phys. Rev.} {\bf D23}, 347 (1981).}
\REF\four{T.W.G. Kibble, {\it Phys. Rep.} {\bf 67}, 183 (1980).}
\REF\five{A. Vilenkin, {\it Phys. Rep.} {\bf 121}, 263 (1985).}
\REF\six{N. Turok, `Phase Transitions as the Origin of Large-Scale
Structure,' in `Particles, Strings and Supernovae' (TASI-88) ed. by
A. Jevicki and C.-I. Tan (World Scientific, Singapore, 1989).}
\REF\seven{A. Vilenkin and E.P.S. Shellard, `Topological Defects and Cosmology'
(Cambridge Univ. Press, Cambridge, 1993).}
\REF\one{R. Brandenberger, {\it Phys. Scripta} {\bf T36}, 114 (1991).}
\REF\eight{N. Turok, {\it Phys. Scripta} {\bf T36}, 135 (1991).}
\REF\eleven{V. de Lapparent, M. Geller and J. Huchra, {\it Ap. J.
(Lett)} {\bf 302}, L1 (1986).}
\REF\one{N. Bahcall and R. Soneira, {\it Ap. J.} {\bf 270}, 20 (1983);
\nextline
A. Klypin and A. Kopylov, {\it Sov. Astr. Lett.} {\bf 9}, 41 (1983).}
\REF\twelve{G. Smoot et al., {\it Ap. J. (Lett.)} {\bf 396}, L1
(1992).}
\REF\one{M. Mermin, {\it Rev. Mod. Phys.} {\bf  51}, 591 (1979).}
\REF\one{P. de Gennes, ``The Physics of Liquid Crystals" (Clarendon Press,
Oxford, 1974).}
\REF\one{I. Chuang, R. Durrer, N. Turok and B. Yurke, {\it Science} {\bf 251},
1336 (1991);\nextline
M. Bowick, L. Chandar, E. Schiff and A. Srivastava, Syracuse preprint
SU-HEP-4241-512 (1992).}
\REF\one{M. Salomaa and G. Volovik, {\it Rev. Mod. Phys.} {\bf 59}, 533
(1987).}
\REF\one{A. Abrikosov, {\it JETP} {\bf 5}, 1174 (1957).}
\REF\one{R. Brandenberger, `Lectures on Modern Cosmology and Structure
Formation', Brown preprint BROWN-HET-893 (1993), to be publ. in the proc. of
the $7^{th}$ Swieca Summer School in Particles and Fields, Jan. 10-23 1993,
eds. O. Eboli and V. Ribelles (World Scientific, Singapore, 1993).}
\REF\nine{R. Brandenberger, in `Physics of the Early Universe,' proc.
of the 1989 Scottish Univ. Summer School in Physics, ed. by J. Peacock, A.
Heavens and A. Davies (SUSSP Publ., Edinburgh, 1990).}
\REF\nineteen{E. Milne, {\it Zeits. f. Astrophys.} {\bf 6}, 1 (1933).}
\REF\twenty{S. Shechtman, P. Schechter, A. Oemler, D. Tucker, R. Kirshner and
H. Lin, Harvard-Smithsonian preprint CFA 3385 (1992),
to appear in `Clusters and Superclusters of Galaxies', ed. by A. Fabian
(Kluwer, Dordrecht, 1993).}
\REF\twentyone{see e.g., S. Weinberg, `Gravitation and Cosmology'
(Wiley, New York, 1972); \nextline
Ya.B. Zel'dovich and I. Novikov, `The Structure and Evolution of the
Universe' (Univ. of Chicago Press, Chicago, 1983).}
\REF\twentytwo{E. Hubble, {\it Proc. Nat. Acad. Sci.} {\bf 15}, 168
(1927).}
\REF\twentythree{R. Alpher and R. Herman, {\it Rev. Mod. Phys.} {\bf
22}, 153 (1950); \nextline
G. Gamov, {\it Phys. Rev.} {\bf 70}, 572 (1946).}
\REF\twentyfour{R. Alpher, H. Bethe and G. Gamov, {\it Phys. Rev.}
{\bf 73}, 803 (1948); \nextline
R. Alpher and R. Herman, {\it Nature} {\bf 162}, 774 (1948).}
\REF\twentyfive{For an excellent introduction see S. Weinberg, `The
First Three Minutes' (Basic Books, New York, 1988).}
\REF\twentysix{R. Dicke, P.J.E. Peebles, P. Roll and D. Wilkinson,
{\it Ap. J.} {\bf 142}, 414 (1965).}
\REF\twentyseven{A. Penzias and R. Wilson, {\it Ap. J.} {\bf 142}, 419
(1965).}
\REF\twentyeight{J. Mather et al., {\it Ap. J. (Lett.)} {\bf 354}, L37
(1990).}
\REF\twentynine{H. Gush, M. Halpern and E. Wishnow, {\it Phys. Rev.
Lett.} {\bf 65}, 937 (1990).}
\REF\thirty{see e.g., G. Efstathiou, in `Physics of the Early
Universe,' proc. of the 1989 Scottish Univ. Summer School in Physics,
ed. by J. Peacock, A. Heavens and A. Davies (SUSSP Publ., Edinburgh,
1990).}
\REF\thirtyone{B. Pagel, {\it Ann. N.Y. Acad. Sci.} {\bf 647}, 131
(1991).}
\REF\thirtytwo{H. Arp, G. Burbidge, F. Hoyle, J. Narlikar and N.
Vickramasinghe, {\it Nature} {\bf 346}, 807 (1990).}
\REF\thirtythree{N. Tsamis and R. Woodard, {\it Phys. Lett.} {\bf 301B}, 351
(1993).}
\REF\thirtyfour{J. Tonry, in `Relativistic Astrophysics and Particle Cosmology'
(TEXAS-PASCOS 92), Berkeley,
Dec. 12-18, 1992, publ.in {\it Ann. N.Y. Academy of Sciences} {\bf 688}, 114
(1993).}
\REF\thirtyfive{V. Trimble, {\it Ann. Rev. Astr. Astrophys.} {\bf 25}, 423
(1987).}
\REF\thirtysix{E. Bertschinger and A. Dekel, {\it Ap. J.} {\bf 336},
L5 (1989).}
\REF\thirtyseven{M. Strauss, M. Davis, A. Yahil and J. Huchra, {\it
Ap. J.} {\bf 385}, 421 (1992).}
\REF\thirtyeight{G. Ellis, {\it Class. Quant. Grav.} {\bf 5}, 207 (1988);
\nextline
G. Ellis, D. Lyth and M. Mijic, {\it Phys. Lett.} {\bf 271B}, 52
(1991);\nextline
M. Madsen, G. Ellis, J. Mimoso and J. Butcher, {\it Phys. Rev.} {\bf D46}, 1399
(1992).}
\REF\thirtynine{J. Primack, D. Seckel and B. Sadoulet, {\it Ann. Rev. Nucl.
Part. Sci.} {\bf 38}, 751 (1988).}
\REF\forty{L. da Costa, in `The Distribution of Matter in the
Universe' ed. by D. Gerbal and G. Mamon, in press (1991).}
\REF\fortyone{T. Broadhurst, R. Ellis, D. Koo and A. Szalay, {\it
Nature} {\bf 343}, 726 (1990).}
\REF\fortytwo{J. Ostriker and L. Cowie, {\it Ap. J. (Lett.)} {\bf
243}, L127 (1981).}
\REF\one{A. Linde, `Particle Physics and Inflationary Cosmology'
(Harwood, Chur, 1990).}
\REF\two{S. Blau and A. Guth, `Inflationary Cosmology,' in `300 Years
of Gravitation' ed. by S. Hawking and W. Israel (Cambridge Univ.
Press, Cambridge, 1987).}
\REF\three{K. Olive, {\it Phys. Rep.} {\bf 190}, 307 (1990).}
\REF\fortythree{D. Kazanas, {\it Ap. J.} {\bf 241}, L59 (1980).}
\REF\fortyfour{W. Press, {\it Phys. Scr.} {\bf 21}, 702 (1980).}
\REF\fortyfive{G. Chibisov and V. Mukhanov, `Galaxy Formation and
Phonons,' Lebedev Physical Institute Preprint No. 162 (1980);
\nextline
G. Chibisov and V. Mukhanov, {\it Mon. Not. R. Astron. Soc.} {\bf
200}, 535 (1982).}
\REF\fortysix{V. Lukash, {\it Pis'ma Zh. Eksp. Teor. Fiz.} {\bf 31}, 631
(1980).}
\REF\fortyseven{K. Sato, {\it Mon. Not. R. Astron. Soc.} {\bf 195},
467 (1981).}
\REF\one{L. Landau and E. Lifshitz, `Theoretical Physics, Vol. II:
Classical Fields' (Pergamon, London, 1958).}
\REF\fortyeight{D. Kirzhnits and A. Linde, {\it Pis'ma Zh. Eksp.
Teor. Fiz.} {\bf 15}, 745 (1972); \nextline
D. Kirzhnits and A. Linde, {\it Zh. Eksp. Teor. Fiz.} {\bf 67}, 1263
(1974);\nextline
C. Bernard, {\it Phys. Rev.} {\bf D9}, 3313 (1974);\nextline
L. Dolan and R. Jackiw, {\it Phys. Rev.} {\bf D9}, 3320 (1974);\nextline
S. Weinberg, {\it Phys. Rev.} {\bf D9}, 3357 (1974).}
\REF\eight{R. Brandenberger, {\it Rev. Mod. Phys.} {\bf 57}, 1
(1985).}
\REF\fourteen{A. Linde, {\it Phys. Lett.} {\bf 129B}, 177 (1983).}
\REF\fortynine{G. Mazenko, W. Unruh and R. Wald, {\it Phys. Rev.} {\bf
D31}, 273 (1985).}
\REF\fifty{A. Linde, {\it Phys. Lett.} {\bf 108B}, 389 (1982);
\nextline
A. Albrecht and P. Steinhardt, {\it Phys. Rev. Lett.} {\bf 48}, 1220
(1982).}
\REF\fone{J. Langer, {\it Physica} {\bf 73}, 61 (1974).}
\REF\ftwo{S. Coleman, {\it Phys. Rev.} {\bf D15}, 2929 (1977);
\nextline
C. Callan and S. Coleman, {\it Phys. Rev.} {\bf D16}, 1762 (1977).}
\REF\fthree{M. Voloshin, Yu. Kobzarev and L. Okun, {\it Sov. J. Nucl.
Phys.} {\bf 20}, 644 (1975).}
\Ref\one{M. Stone, {\it Phys. Rev.} {\bf D14}, 3568 (1976);\nextline
M. Stone, {\it Phys. Lett.} {\bf 67B}, 186 (1977).}
\Ref\one{P. Frampton, {\it Phys. Rev. Lett.}, {\bf 37}, 1380 (1976).}
\Ref\one{S. Coleman, in `The Whys of Subnuclear Physics' (Erice 1977), ed by
A. Zichichi (Plenum, New York, 1979).}
\REF\one{G. Ross, Grand Unified Theories (Benjamin, Reading,
1985).}
\REF\one{S. Coleman, `Secret Symmetry' in `New Phenomena in Subnuclear Physics'
(Erice 1975), ed. by A. Zichichi
(Plenum Press, New York, 1977).}
\REF\one{A. Starobinsky, {\it Phys. Lett.} {\bf 91B}, 99 (1980).}
\REF\one{N. Steenrod, `Topology of Fibre Bundles' (Princeton Univ. Press,
Princeton, 1951).}
\REF\one{H. Nielsen and P. Olesen, {\it Nucl. Phys.} {\bf B61}, 45
(1973).}
\REF\one{R. Davis, {\it Phys. Rev.} {\bf D35}, 3705 (1987).}
\REF\one{N. Turok, {\it Phys. Rev. Lett.} {\bf 63}, 2625 (1989).}
\REF\one{P. Langacker and S.-Y. Pi, {\it Phys. Rev. Lett.} {\bf 45}, 1
(1980).}
\REF\one{T.W.B. Kibble and E. Weinberg, {\it Phys. Rev.} {\bf D43},
3188 (1991).}
\REF\one{T.W.B. Kibble, {\it Acta Physica Polonica} {\bf B13}, 723
(1982).}
\REF\one{S. Rudaz and A. Srivastava, `On the Production of Flux
Vortices and Magnetic Monopoles in Phase Transitions,' Univ. of
Minnesota preprint UMN-TH-1028/92 (1992).}
\REF\one{J. Ye and R. Brandenberger, {\it Nucl. Phys.} {\bf B346}, 149
(1990).}
\REF\one{M. Hindmarsh, A.-C. Davis and R. Brandenberger, `Formation of
Topological Defects in First Order Phase Transitions,' Brown Univ.
preprint BROWN-HET-902 (1993).}
\REF\one{Ya.B. Zel'dovich, I. Kobzarev and L. Okun, {\it Zh. Eksp.
Teor. Fiz.} {\bf 67}, 3 (1974).}
\REF\one{Ya.B. Zel'dovich and M. Khlopov, {\it Phys. Lett.} {\bf 79B},
239 (1978); \nextline
J. Preskill, {\it Phys. Rev. Lett.} {\bf 43}, 1365 (1979).}
\REF\one{M. Barriola and A. Vilenkin, {\it Phys. Rev. Lett.} {\bf 63},
341 (1989); \nextline
S. Rhie and D. Bennett, {\it Phys. Rev. Lett.} {\bf 65}, 1709
(1990).}
\REF\one{T. Vachaspati and A. Vilenkin, {\it Phys. Rev.} {\bf D30},
2036 (1984).}
\REF\one{Ya.B. Zel'dovich, {\it Mon. Not. R. astron. Soc.} {\bf 192},
663 (1980);\nextline
A. Vilenkin, {\it Phys. Rev. Lett.} {\bf 46}, 1169 (1981).}
\REF\one{D. Foerster, {\it Nucl. Phys.} {\bf B81}, 84 (1974).}
\REF\one{N. Turok, in `Proceedings of the 1987 CERN/ESO Winter School
on Cosmology and Particle Physics' (World Scientific, Singapore,
1988).}
\REF\one{T.W.B. Kibble and N. Turok, {\it Phys. Lett.} {\bf 116B}, 141
(1982).}
\REF\one{R. Brandenberger, {\it Nucl. Phys.} {\bf B293}, 812 (1987).}
\REF\one{E.P.S. Shellard, {\it Nucl. Phys.} {\bf B283}, 624 (1987).}
\REF\one{R. Matzner, {\it Computers in Physics} {\bf 1}, 51 (1988);
\nextline
K. Moriarty, E. Myers and C. Rebbi, {\it Phys. Lett.} {\bf 207B}, 411
(1988); \nextline
E.P.S. Shellard and P. Ruback, {\it Phys. Lett.} {\bf 209B}, 262
(1988).}
\REF\one{P. Ruback, {\it Nucl. Phys.} {\bf B296}, 669 (1988).}
\REF\one{T. Vachaspati and A. Vilenkin, {\it Phys. Rev.} {\bf D31},
3052 (1985); \nextline
N. Turok, {\it Nucl. Phys.} {\bf B242}, 520 (1984); \nextline
C. Burden, {\it Phys. Lett.} {\bf 164B}, 277 (1985).}
\REF\one{A. Albrecht and N. Turok, {\it Phys. Rev. Lett.} {\bf 54},
1868 (1985).}
\REF\one{D. Bennett and F. Bouchet, {\it Phys. Rev. Lett.} {\bf 60},
257 (1988).}
\REF\one{B. Allen and E.P.S. Shellard, {\it Phys. Rev. Lett.} {\bf
64}, 119 (1990).}
\REF\one{A. Albrecht and N. Turok, {\it Phys. Rev. } {\bf D40}, 973
(1989).}
\REF\one{R. Brandenberger and J. Kung, in `The Formation and Evolution
of Cosmic Strings', eds. G. Gibbons, S. Hawking and T. Vachaspati
(Cambridge Univ. Press, Cambridge, 1990).}
\REF\one{see e.g., C. Misner, K. Thorne and J. Wheeler, `Gravitation'
(Freeman, San Francisco, 1973).}
\REF\one{B. Carter, {\it Phys. Rev.} {\bf D41}, 3869 (1990).}
\REF\one{E. Copeland, T.W.B. Kibble and D. Austin, {\it Phys. Rev.}
{\bf D45}, 1000 (1992).}
\REF\one{T. Prokopec, {\it Phys. Lett.} {\bf 262B}, 215 (1991);\nextline
R. Leese and T. Prokopec, {\it Phys. Rev.} {\bf D44}, 3749
(1991).}
\REF\one{D. Spergel, N. Turok, W. Press and B. Ryden, {\it Phys.Rev.} {\bf
D43}, 1038 (1991).}
\REF\one{T. Prokopec, A. Sornborger and R. Brandenberger, {\it Phys.
Rev.} {\bf D45}, 1971 (1992).}
\REF\one{J. Borrill, E. Copeland and A. Liddle, {\it Phys. Lett.} {\bf 258B},
310 (1991).}
\REF\one{A. Sornborger, `A Semi-Analytic Study of Texture
Collapse,' Brown Univ. preprint BROWN-HET-895 (1993), {\it Phys. Rev.} {\bf D},
in press.}
\REF\one{L. Perivolaropoulos, {\it Phys. Rev.} {\bf D46}, 1858
(1992).}
\REF\one{E. Lifschitz, {\it Zh. Ekop. Teor. Fiz.} {\bf 16}. 587 (1946).}
\REF\fifteen{V. Mukhanov, H. Feldman and R. Brandenberger, {\it Phys.
Rep.} {\bf 215}, 203 (1992).}
\REF\onefortyeight{L. Krauss and M. White, {\it Phys. Rev. Lett.} {\bf 69}, 869
(1992);\nextline
D. Salopek, {\it Phys. Rev. Lett.} {\bf 69}, 3602 (1992);\nextline
R. Davis, H. Hodges, G. Smoot, P. Steinhardt and M. Turner, {\it Phys. Rev.
Lett.} {\bf 69}, 1856 (1992);\nextline
A. Liddle and D. Lyth, {\it Phys. Lett.} {\bf 291B}, 391 (1992); \nextline
F. Lucchin, S. Matarrese and S. Mollerach, {\it Ap. J. (Lett.)}, {\bf 401}, L49
(1992);\nextline
T. Souradeep and V. Sahni, {\it Mod. Phys. Lett.} {\bf A7}, 3541 (1992).}
\REF\one{D. Bennett, A. Stebbins and F. Bouchet, {\it Ap. J. (Lett.)}
{\bf 399}, L5 (1992).}
\REF\one{D. Bennett and S. Rhie,  {\it Ap. J. (Lett.)} {\bf 406}, L7 (1993).}
\REF\one{L. Perivolaropoulos, {\it Phys. Lett.} {\bf 298B}, 305 (1993).}
\REF\one{R. Sachs and A. Wolfe, {\it Ap. J.} {\bf 147}, 73 (1967).}
\REF\one{M. Rees and D. Sciama, {\it Nature} {\bf 217}, 511 (1968).}
\REF\one{D. Fixsen, E. Cheng and D. Wilkinson, {\it Phys. Rev. Lett} {\bf 50},
620 (1983).}
\REF\one{K. Ganga, E. Cheng, S. Meyer and L. Page, {\it Ap. J. (Lett.)} {\bf
410}, L54 (1993).}
\REF\one{A. Lasenby,  `Ground-Based Observations of the Cosmic Microwave
Background', Cambridge Univ. preprint (1992), to be publ. in the proc. of the
2nd Course of the International School of Astrophysics `D. Chalonge',
6 - 13 Sept. 1992, Erice, Italy, ed. N. Sanchez (World Scientific, Singapore,
1993).}
\REF\one{P. Coles and J. Barrow, {\it Mon. Not. R. astron. Soc.} {\bf 228}, 407
(1987);\nextline
P. Coles, {\it Mon. Not. R. astron. Soc.} {\bf 234}, 501 (1988);\nextline
J. Gott et al., {\it Ap. J.} {\bf 340}, 625 (1989).}
\REF\one{L. Perivolaropoulos, `On the Statistics of CMB Fluctuations induced by
Topological Defects', Harvard-Smithsonian Center for
 Astrophysics preprint
CFA 3526 (1992); \nextline
R. Moessner, L. Perivolaropoulos and R. Brandenberger, `A Cosmic String
Specific Signature on the Cosmic Microwave Background', Brown Univ. preprint
BROWN-HET-911 (1993), Ap. J. (in press);\nextline
P. Graham, N. Turok, P. Lubin and J. Schuster, `A Simple Test for
Non-Gaussianity in CMBR Measurements', Princeton preprint PUP-TH-1408 (1993).}
\REF\one{Ya.B. Zel'dovich, J. Einasto and S. Shandarin, {\it Nature}
{\bf 300}, 407 (1982); \nextline
J. Oort, {\it Ann. Rev. Astron. Astrophys.} {\bf 21}, 373 (1983);
\nextline
R.B. Tully, {\it Ap. J.} {\bf 257}, 389 (1982); \nextline
S. Gregory, L. Thomson and W. Tifft, {\it Ap. J.} {\bf 243}, 411
(1980).}
\REF\one{G. Chincarini and H. Rood, {\it Nature} {\bf 257}, 294
(1975); \nextline
J. Einasto, M. Joeveer and E. Saar, {\it Mon. Not. R. astron. Soc.}
{\bf 193}, 353 (1980); \nextline
R. Giovanelli and M. Haynes, {\it Astron. J.} {\bf 87}, 1355 (1982);
\nextline
D. Batuski and J. Burns, {\it Ap. J.} {\bf 299}, 5 (1985).}
\REF\one{R. Kirshner, A. Oemler, P. Schechter and S. Shechtman, {\it
Ap. J. (Lett.)} {\bf 248}, L57 (1981).}
\REF\one{M. Joeveer, J. Einasto and E. Tago, {\it Mon. Not. R. astron.
Soc.} {\bf 185}, 357 (1978); \nextline
L. da Costa et al., {\it Ap. J.} {\bf 327}, 544 (1988).}
\REF\one{A. Dressler, D. Lynden-Bell, D. Burstein, R. Davies, S.
Faber, R. Terlevich and G. Wegner, {\it Ap. J.} {\bf 313}, 42
(1987);\nextline
C. Collins, R. Joseph and N. Robertson, {\it Nature} {\bf 320}, 506
(1986).}
\REF\one{E. Bertschinger and A. Dekel, {\it Ap. J. (Lett.)} {\bf 336}, L5
(1989).}
\REF\one{M. Aaronson, G. Bothun, J. Mould, R. Schommer and M. Cornell,
{\it Ap. J.} {\bf 302}, 536 (1986); \nextline
J. Lucey and D. Carter, {\it Mon. Not. R. astron. Soc.} {\bf 235},
1177 (1988).}
\REF\one{G. Abell, {\it Ap. J. Suppl.} {\bf 3}, 211 (1958).}
\REF\one{M. Davis and J. Huchra, {\it Ap. J.} {\bf 254}, 437 (1982).}
\REF\one{P.J.E. Peebles, in `Physical Cosmology,' 1979 Les Houches
Lectures, ed. by R. Balian, J. Audouze and D. Schramm (North-Holland,
Amsterdam, 1980).}
\REF\one{A. Dekel, G. Blumenthal, J. Primack and S. Olivier, {\it Ap.
J. (Lett.)} {\bf 338}, L5 (1989); \nextline
W. Sutherland, {\it Mon. Not. R. astron. Soc.} {\bf 234}, 159 (1988).}
\REF\one{N. Bahcall, {\it Ann. Rev. Astr. \& Astrophys.} {\bf 15}, 505
(1977); \nextline
B. Binggeli, in `Nearly Normal Galaxies,' ed. S. Faber (Springer, New
York, 1986).}
\REF\one{see e.g., E. Athanassoula and A. Bosma, in `Large Scale
Structures of the Universe,' IAU Symposium No. 130, ed. by J. Audouze
et al. (Kluwer, Dordrecht, 1988).}
\REF\one{see e.g., S.M. Fall, in `Internal Kinetics and Dynmaics of
Galaxies,' ed. by E. Athanassoula (Reidel, Dordrecht, 1983).}
\REF\one{N. Turok, {\it Nucl. Phys.} {\bf B242}, 520 (1984).}
\REF\one{N. Turok and R. Brandenberger, {\it Phys. Rev.} {\bf D33},
2175 (1986); \nextline
A. Stebbins, {\it Ap. J. (Lett.)} {\bf 303}, L21 (1986); \nextline
H. Sato, {\it Prog. Theor. Phys.} {\bf 75}, 1342 (1986).}
\REF\one{A. Vilenkin, {\it Phys. Rev.} {\bf D23}, 852 (1981);
\nextline
J. Gott, {\it Ap. J.} {\bf 288}, 422 (1985); \nextline
W. Hiscock, {\it Phys. Rev.} {\bf D31}, 3288 (1985); \nextline
B. Linet, {\it Gen. Rel. Grav.} {\bf 17}, 1109 (1985); \nextline
D. Garfinkle, {\it Phys. Rev.} {\bf D32}, 1323 (1985); \nextline
R. Gregory, {\it Phys. Rev. Lett.} {\bf 59}, 740 (1987).}
\REF\one{J. Silk and V. Vilenkin, {\it Phys. Rev. Lett.} {\bf 53},
1700 (1984).}
\REF\one{T. Vachaspati, {\it Phys. Rev. Lett.} {\bf 57}, 1655 (1986).}
\REF\one{A. Stebbins, S. Veeraraghavan, R. Brandenberger, J. Silk and
N. Turok, {\it Ap. J.} {\bf 322}, 1 (1987).}
\REF\one{R. Brandenberger, L. Perivolaropoulos and A. Stebbins, {\it
Int. J. of Mod. Phys.} {\bf A5}, 1633 (1990); \nextline
L. Perivolarapoulos, R. Brandenberger and A. Stebbins, {\it Phys.
Rev.} {\bf D41}, 1764 (1990).}
\REF\one{E. Bertschinger and P. Watts, {\it Ap. J.} {\bf 328}, 23 (1988).}
\REF\one{D. Vollick, {\it Phys. Rev.} {\bf D45}, 1884 (1992).}
\REF\one{T. Vachaspati and A. Vilenkin, {\it Phys. Rev. Lett.} {\bf 67}, 1057
(1991).}
\REF\one{A. Gooding, D. Spergel and N. Turok, {\it Ap. J. (Lett.)}
{\bf 372}, L5 (1991).}
\REF\one{C. Park, D. Spergel and N. Turok, {\it Ap. J. (Lett.)} {\bf 373}, L53
(1991).}
\REF\one{R. Cen, J. Ostriker, D. Spergel and N. Turok, {\it Ap. J.}
{\bf 383}, 1 (1991).}
\REF\one{A. Gooding, C. Park, D. Spergel, N. Turok and J. Gott, {\it
Ap. J.} {\bf 393}, 42 (1992).}
\REF\one{L. Grishchuk and Y. Sidorov, {\it Class. Quant. Grav.} {\bf
6}, L161 (1989); \nextline
L. Grishchuk and Y. Sidorov, {\it Phys. Rev.} {\bf D42}, 3413 (1990);
\nextline
L. Grishchuk, {\it Phys. Rev. Lett.} {\bf 70}, 2371 (1993).}
\REF\one{T. Prokopec, `Entropy of the Squeezed Vacuum,' Brown Univ.
preprint BROWN-HET-861 (1992); \nextline
A. Albrecht, P. Ferreira, M. Joyce and T. Prokopec, `Inflation and
Squeezed Quantum States,' Imperial College preprint (1993).}
\REF\one{R. Brandenberger, N. Kaiser, D. Schramm and N. Turok, {\it
Phys. Rev. Lett.} {\bf 59}, 2371 (1987).}
\REF\one{R. Brandenberger, N. Kaiser and N. Turok, {\it Phys. Rev.}
{\bf D36}, 2242 (1987).}
\REF\one{J. Gott, A. Melott and M. Dickinson, {\it Ap. J.} {\bf 306},
341 (1986).}
\REF\one{J. Gerber and R. Brandenberger, `Topology of Large-Scale
Structure in a Cosmic String Wake Model,' Brown Univ. preprint
BROWN-HET-829 (1991).}
\REF\one{W. Saslaw, {\it Ap. J.} {\bf 297}, 49 (1985).}
\REF\one{S. Ramsey, Senior thesis, Brown Univ. (1992); \nextline
D. M. Kaplan, Senior thesis, Brown Univ. (1993);\nextline
R. Brandenberger, D.M. Kaplan and S. Ramsey, `Some Statistics for Measuring
Large-Scale Structure', Brown preprint BROWN-HET-922 (1993).}
\REF\one{E. Valentini and R. Brandenberger, in preparation (1993);
\nextline
D. Weinberg and S. Cole, {\it Mon. Not. R. astron. Soc.} {\bf 259}, 652 (1992)
.}
\REF\one{S. White, C. Frenk, M. Davis and G. Efstathiou, {\it Ap. J.}
{\bf 313}, 505 (1987).}
\REF\one{N. Kaiser and A. Stebbins, {\it Nature} {\bf 310}, 391
(1984).}
\REF\one{J. Traschen, N. Turok and R. Brandenberger, {\it Phys. Rev.}
{\bf D34}, 919 (1986); \nextline
S. Veeraraghavan and A. Stebbins, {\it Ap. J. (Lett.)} {\bf 395}, L55 (1992).}
\REF\one{D. Bennett and F. Bouchet, {\it Phys. Rev.} {\bf D41}, 2408 (1990).}
\REF\one{R. Durrer and D. Spergel, `Microwave Anisotropies from
Texture Seeded Structure Formation,' Princeton preprint PUPT-91-1247
(1991);\nextline
R. Durrer, D. Spergel and A. Hayward, `Microwave Anisotropies from Texture
Seeded Structure Formation', Z\"urich Univ. preprint ZH-TH32/92 (1992).}
\REF\oneninetyseven{N. Turok and D. Spergel, {\it Phys. Rev. Lett.} {\bf 64},
2736 (1990).}
\REF\one{U.-L. Pen, D. Spergel and N. Turok, `Cosmic Structure Formation and
Microwave Anisotropies from Global Field Ordering, Princeton preprint
PU-TH-1375 (1993).}
\REF\one{C. Callan, {\it Phys.  Rev.} {\bf D25}, 2141 (1982);
{\it Phys. Rev.} {\bf D26}, 2058 (1982).}
\REF\one{V. Rubakov, {\it Pis'ma Zh. Eksp. Fiz.} {\bf 33}, 658 (1981) [JETP
Lett. 33, 644 (1981)];
{\it Nucl. Phys.} {\bf B203}, 311 (1982).}
\REF\one{R. Brandenberger, A.-C. Davis and A. Matheson, {\it Nucl. Phys.} {\bf
B307}, 909 (1988).}
\REF\one{R. Brandenberger and L. Perivolaropoulos, {\it Phys. Lett.} {\bf
B208}, 396 (1988).}
\REF\one{Y. Aharanov and D. Bohm, {\it Phys. Rev.} {\bf 119}, 485 (1959).}.
\REF\one{M. Alford and F. Wilczek, {\it Phys. Rev. Lett.} {\bf 62}, 1071
(1989).}
\REF\one{P. de Sousa Gerbert, {\it Phys. Rev.} {\bf D40}, 1346 (1989);
\nextline
P. de Sousa Gerbert and R. Jackiw, {\it Comm. Math. Phys.} {\bf 124}, 229
(1989).}
\REF\one{M. Alford, J. March-Russell and F. Wilczek, {\it Nucl. Phys.} {\bf
B328}, 140 (1989).}
\REF\one{W. Perkins, L. Perivolaropoulos, A.-C. Davis, R. Brandenberger and A.
Matheson,
{\it Nucl. Phys.} {\bf B353}, 237 (1991).}
\REF\one{H. de Vega, {\it Phys. Rev.} {\bf D18}, 2932 (1978).}
\REF\one{L. Perivolaropoulos, A. Matheson, A.-C. Davis and R. Brandenberger,
{\it
Phys. Lett.} {\bf B245}, 556 (1990).}
\REF\one{T.W.B. Kibble, G. Lazarides and Q. Shafi, {\it Phys. Lett.} {\bf
B113}, 237 (1982).}
\REF\one{D. Olive and N. Turok, {\it Phys. Lett.} {\bf B117}, 193 (1982).}
\REF\one{M. Hindmarsh and T.W.B. Kibblem, {\it Phys. Rev. Lett.} {\bf 55}, 2398
(1985).}
\REF\one{M. Aryal and A. Everett, {\it Phys. Rev.} {\bf D35}, 3105 (1987).}
\REF\one{T.T. Wu and C.N. Yang, {\it Phys. Rev.} {\bf D12}, 3845 (1975).}
\REF\one{P. Horvathy, {\it Phys. Rev.} {\bf D33}, 407 (1985).}
\REF\one{C.-P. Ma, {\it Phys. Rev.} {\bf D48}, 530 (1993).}
\REF\one{A. Sakharov, {\it Zh. Eksp. Teor. Fiz. Pis'ma} {\bf 5}, 32 (1967).}
\REF\one{M. Yoshimura, {\it Phys. Rev. Lett.} {\bf 41}, 281 (1978); \nextline
A. Ignatiev, N. Krasnikov, V. Kuzmin and A. Tavkhelidze, {\it Phys. Lett.} {\bf
B76}, 436 (1978);
\nextline
S. Dimopoulos and L. Susskind, {\it Phys. Rev.} {\bf  D18}, 4500 (1978);
\nextline
S. Weinberg, {\it Phys. Rev. Lett.} {\bf 42}, 850 (1979); \nextline
D. Toussaint, S. Trieman, F. Wilczek and A. Zee, {\it Phys. Rev.} {\bf D19},
1036 (1979).}
\REF\one{see e.g. S. Weinberg, 'Gravitation and Cosmology' (Wiley, New York,
1972).}
\REF\one{P. Bhattacharjee, T.W.B. Kibble and N. Turok, {\it Phys. Lett.} {\bf
B119}, 95 (1982).}
\REF\one{M. Kawasaki and K. Maeda, {\it Phys. Lett.} {\bf B209}, 271 (1988).}
\REF\one{R. Brandenberger, {\it Nucl. Phys.} {\bf B293}, 812 (1987).}
\REF\one{R. Brandenberger, A.-C. Davis and M. Hindmarsh, {\it Phys. Lett.} {\bf
263}, 239 (1991).}
\REF\one{D. Nanopoulos and S. Weinberg, {\it Phys. Rev.} {\bf D20}, 2484
(1979).}
\REF\one{G. 't Hooft, {\it Phys. Rev. Lett.} {\bf 37}, 8 (1976).}
\REF\one{L. Ib\'a\~nez and F. Quevedo, {\it Phys. Lett.} {\bf B283}, 261
(1992).}
\REF\one{M. Shaposhnikov, {\it JETP Lett.} {\bf 44}, 465 (1986); \nextline
M. Shaposhnikov, {\it Nucl. Phys.} {\bf B287}, 757 (1987); \nextline
L. McLerran, {\it Phys. Rev. Lett.} {\bf 62}, 1075 (1989).}
\REF\one{N. Turok and J. Zadrozny, {\it Phys. Rev. Lett.} {\bf 65}, 2331
(1990);\nextline
N. Turok and J. Zadrozny, {\it Nucl. Phys} {\bf B358}, 471 (1991); \nextline
L. McLerran, M. Shaposhnikov, N. Turok and M. Voloshin, {\it Phys. Lett.} {\bf
B256}, 451 (1991).}
\REF\one{A. Cohen, D. Kaplan and A. Nelson, {\it Phys. Lett.} {\bf B263}, 86
(1991).}
\REF\one{A. Nelson, D. Kaplan and A. Cohen, {\it Nucl. Phys.} {\bf B373}, 453
(1992).}
\REF\one{R. Brandenberger and A.-C. Davis, {\it Phys. Lett.} {\bf B308}, 79
(1993).}
\REF\one{T. Vachaspati, {\it Phys. Rev. Lett.} {\bf 68}, 1977 (1992).}
\REF\one{Y. Nambu, {\it Nucl. Phys.} {\bf B130}, 505 (1977).}
\REF\one{M. James, L. Perivolaropoulos and T. Vachaspati, {\it Phys.Rev} {\bf
D46}, R5232 (1992).}
\refout

\end